\documentclass[structabstract]{aa}  
\usepackage{graphicx}
\usepackage{color}
\usepackage{url}
\usepackage{amssymb}
\usepackage{amsmath}
\usepackage{txfonts}
\usepackage{multirow}
\usepackage{lscape}
\usepackage[flushleft]{threeparttable}
\usepackage{float}
\usepackage{natbib}
\usepackage{stmaryrd}
\usepackage{supertabular}
\usepackage{rotating}
\usepackage{longtable}
\usepackage{chngcntr}
\bibpunct{(}{)}{;}{a}{}{,} % to follow the A&A style

\makeatletter
\def\hlinewd#1{%
\noalign{\ifnum0=`}\fi\hrule \@height #1 %
\futurelet\reserved@a\@xhline}
\makeatother

\newfont{\gwpfont}{cmssq8 scaled 1000}

\begin{document}
   \title{SARCS strong-lensing galaxy groups:\\
    II - mass-concentration relation and strong-lensing bias
    \thanks{Strong Lensing Legacy Survey SL2S-ARCS}}

   \author{
          G. Fo\"ex\inst{1}
          \and
          V. Motta\inst{1}
          \and
          E. Jullo\inst{2}
          \and
          M. Limousin\inst{2,3}
          \and
          T. Verdugo\inst{4}
          }
   \institute{
          Instituto de F\'isica y Astronom\'ia, Universidad de Valpara\'iso, Avda. Gran Breta\~{n}a 1111, Valpara\'iso, Chile
          \and
          Aix Marseille Universit\'e, CNRS, LAM (Laboratoire d'Astrophysique de Marseille) UMR 7326, 13388, Marseille, France
          \and
          Dark Cosmology Centre, Niels Bohr Institute, University of Copenhagen, Juliane Maries Vej 30, DK-2100 Copenhagen, Denmark
          \and
          Centro de Investigaciones de Astronom\'ia, AP 264, M\'erida 5101-A, Venezuela
             }

   \date{Received ; accepted }

  \abstract  
 {} 
  {Various studies have shown a lensing bias in the mass-concentration relation of cluster-scale structures due to alignment of the major axis and the line of sight. In this paper, we aim to study this lensing bias through the mass-concentration relation of galaxy groups, thus extending observational constraints to dark matter haloes of mass $\sim10^{13}-10^{14}\,\mathrm{M_{\odot}}$.}
  {Our work is based on the stacked weak-lensing analysis of a sample of 80 strong-lensing galaxy groups. By combining several lenses, we increase significantly the signal-to-noise ratio of the lensing signal, thus providing constraints on the mass profile that cannot be obtained for individual objects. The resulting shear profiles are fitted with various mass models, among them the NFW profile, which provides an estimate of the total mass and concentration of the composite galaxy groups.}
  {The main results of our analysis are the following: (i) the lensing signal does not allow us to firmly reject a simple singular isothermal sphere mass distribution compared to the expected NFW mass profile; (ii) we obtain an average concentration $c_{200}=8.6_{-1.3}^{+2.1}$ that is much higher than the expected value from numerical simulations for the corresponding average mass $M_{200}=0.73_{-0.10}^{+0.11}\times10^{14}\mathrm{M_{\odot}}$; (iii) the combination of our results with those at larger mass scales gives a mass-concentration relation $c(M)$ over nearly two decades in mass, with a slope in disagreement with predictions from numerical simulations using unbiased populations of dark matter haloes; (iv) our combined $c(M)$ relation matches results from simulations using only haloes with a large strong-lensing cross section, i.e. elongated with a major axis close to the line of sight; (v) for the simplest case of prolate haloes, we estimate with a toy model a lower limit on the minor:major axis ratio $a/c=0.5$ for the average SARCS galaxy group.}
  {Our analysis based on galaxy groups confirmed the results obtained at larger mass scales: strong lenses present apparently too large concentrations, which can be explained by triaxial haloes preferentially oriented with the line of sight. Because more massive systems already have large lensing cross sections, they do not require a large elongation along the line of sight, contrary to less massive galaxy groups. Therefore, it is natural to observe larger lensing (projected) concentrations for such systems, resulting in an overall mass-concentration relation steeper than that of non-lensing haloes.}


   \maketitle

%
%________________________________________________________________

\section{Introduction}

Among the challenges faced by modern cosmology, characterizing the mass distribution of groups and clusters of galaxies arouses a lot of interest, both from the observational and theoretical points of view. These objects, whose mass is thought to be dominated by the so-called dark matter, are theoretically forming in a hierarchical bottom-up fashion when considering only gravitational interactions \citep{kaiser86,white91}. Because the only scale of the process is the mass contained in the initial over-density that leads to the formation of a virialized halo, groups and clusters are supposed to constitute a population of self-similar objects. This 'universality' has been the subject of intensive work, for instance through the theoretical scaling relations between the clusters' total mass and their other observable physical quantities (see e.g. \citealt{giodini13} for a recent review). The analysis of these scaling laws is of prime importance to understand the physics involved in the formation and evolution of structures, in particular with evidence of similarity breaks due to baryonic processes (e.g. \citealt{voit05}).\\
\indent Another way to test and constrain the model of structure formation consists in studying their internal mass distribution. For instance, the three-dimensional shape of the haloes, their average mass profile in the inner parts and at large scales, the influence of the central galaxy and its stellar mass, or the level of substructures resulting from the accretion history are some key aspects of the problem, reflecting the properties of dark matter coupled to the cosmological evolution of the Universe (see e.g. the review by \citealt{bartelmann13} and references therein). Some general predictions can be made from theoretical modeling with simplifying hypotheses, such as the profile of an isothermal mass distribution \citep{binney87}, or the mass density contrast of a virialized halo in the case of spherical collapse in a matter-dominated Universe \citep{gunn72}. However, owing to the complexity of the problem, one has to rely on numerical simulations to derive some statistical properties of the halo population. Over the past two decades, tremendous efforts have been made in this direction, with the emergence of a rather well-defined model of dark matter halo formation within the framework of the $\Lambda$-CDM concordance cosmological model. Despite some intrinsic limitations of such simulations (e.g. spatial and mass resolution, treatment of baryonic physics, properties of the dark matter, etc), they provided over the years several expectations about the mass distribution of group- and cluster-scale haloes. In particular, the dark matter N-body numerical simulations performed by \cite{navarro95,navarro96,navarro97} led to the prediction of a universal profile, able to recover the mass distribution of simulated haloes over three decades in mass. This Navarro-Frenk-White profile (NFW hereafter) is characterized by a rather flat central density and a steeper profile with a logarithmic slope of -3 at larger scales, the transition between the two asymptotical regimes occurring at a so-called scale radius. The properties of this profile and its ability to describe both real and simulated haloes have been intensively explored. Special regards have been given to the so-called concentration parameter, the ratio between the Virial radius and the scale radius. Intuitively, the Universe's background density sets the characteristic density contrast of a halo at its formation epoch. In the case of the NFW model, this density contrast is related to the concentration parameter (see Section 4). Therefore, in a hierarchical scenario where more massive haloes form later, when the background density is smaller, one expects clusters to be less concentrated than groups. This mass-concentration relation has been largely studied in numerical simulations, with, indeed, a bottom-up formation of structures leading to haloes less concentrated at larger mass scales (e.g. \citealt{navarro97,bullock01,eke01,dolag04,neto07,duffy08,gao08,zhao09}). Interestingly, the recent work by \cite{prada12} resulted in a different relation, with a larger normalization and increasing concentrations for more massive galaxy clusters (an 'upturn' in concentrations reported first by \citealt{klypin11}). On the one hand, \cite{meneghetti13} showed that the concentrations measured by \cite{prada12} are biased high compared to those of \cite{duffy08} because of the haloes' selection and binning (maximum circular velocity versus mass), along with a different methodology to estimate the concentrations (means of velocity ratio versus fitting of the spherically averaged mass profile). On the other hand, \cite{ludlow12} explained the upturn in concentrations by the dynamical state of the haloes. They found that the upturn disappears when selecting only relaxed clusters. Systems that are collapsing at the time they are identified in the simulations present a compact configuration: a large fraction of the newly-accreted mass is located at their pericentre, which results in larger concentrations. These apparent discrepancies highlight the importance of controlling the selection function and the estimator of the haloes' parameters, two key aspects to be accounted for when comparing results from different analyses.\\
\indent The mass-concentration relations derived in numerical simulations are very sensitive to the input cosmological parameters (e.g. \citealt{bullock01,neto07,maccio08,dutton14,ludlow14}), so they provide a powerful tool to test the $\Lambda$-CDM cosmological model and the scenario of structure formation. Numerous observational studies, based on X-ray observations (e.g. \citealt{pointecouteau05,vikhlinin06,voigt06,zhang06,buote07,gastaldello07,schmidt07,ettori10,ettori11}) or galaxy kinematic data (e.g. \citealt{rines06,wojtak10}), investigated the mass profile of galaxy clusters, successfully determining concentrations matching those from numerical simulations. On the other hand, most of the lensing-based analyses of galaxy clusters found haloes over-concentrated compared to $\Lambda$-CDM predictions (e.g. \citealt{broadhurst05,comerford07,broadhurst08,umetsu08,oguri09,umetsu10,zitrin10,umetsu11,oguri12,okabe13}). Only few lensing studies found an agreement with the predicted mass-concentration relation (e.g. \citealt{halkola06,limousin08,okabe10a,merten14,umetsu14}). This 'overconcentration problem' led to a questioning of the $\Lambda$-CDM model and its scenario of structure formation, in which extremely high concentrations, or similarly large Einstein radii, are statistically very unlikely (e.g. \citealt{broadhurst08b,zitrin09,zitrin11,meneghetti11}; see, however, \citealt{redlich12}). Several explanations can be invoked to overcome such discrepancies. First, dark matter haloes are not spherical, as seen from elliptical projected observational probes, and in numerical simulations where haloes have triaxial shapes with a preference for prolateness (see e.g. the review by \citealt{limousin12} and references therein). Because most of the observational studies make use of the spherical symmetry assumption, it is straightforward to imagine the impact of projection effects for a highly elongated mass distribution aligned with the line of sight: an enhancement of the projected mass density that leads to apparent over-concentrated haloes. On the other hand, simulated haloes being treated in three dimensions, using or not the same hypothesis of spherical symmetry does not dramatically change the estimated masses and concentrations. Such projections effects, which are inherent in any lensing reconstructions, have been widely studied, and provided a solid way to reconcile observational results with theoretical predictions. For instance, \cite{morandi11a,morandi11b} performed a joint X-ray+lensing analysis to directly constrain the three-dimensional shape of the galaxy cluster A1689, for which several lensing-based studies derived very high concentrations (e.g. \citealt{halkola06,medezinski07}). In doing so, they obtained a highly elongated mass distribution with a concentration compatible with the $\Lambda$-CDM predictions (see also \citealt{corless09,sereno11,sereno12} for a different approach to treat projections effects, based on a bayesian modeling with priors on the haloes' elongation derived from numerical simulations).\\
\indent Because elongated haloes with a major axis close to the line of sight have an increase of their central surface mass density, they have a larger chance to produce a strong-lensing signal. This simple consideration leads to a natural explanation for the systematically large concentrations derived for strong-lensing clusters: a coupling of projection effects with a selection bias in the orientation of haloes. The physical properties of strong lenses have been the subject of several studies based on numerical simulations and targeting haloes with a large strong-lensing cross section (e.g. \citealt{hennawi07,corless07,oguri09,meneghetti10,giocoli14,meneghetti14}). These works highlighted that selecting objects with a large strong-lensing cross section introduces, indeed, an orientation bias in the population of haloes, resulting in larger concentrations derived form their surface mass density. They also revealed the existence of a bias in the intrinsic (3D) concentration of strong-lensing haloes, although the main enhancement of 2D concentrations comes from the projection of elongated mass distributions. By treating these simulated haloes as they would be from lensing observations, they derived specific mass-concentration relations characterized by steeper slopes and higher normalizations, leading to predicted concentrations in good agreement with observational results of strong-lensing galaxy clusters.\\
\indent This 'strong-lensing bias' (projection effects of prolate haloes having a major axis close to the line of sight) gives a natural explanation for over-concentrated haloes. However, the theoretical predictions derived from numerical simulations can also be adjusted in some way. In particular, including baryonic physics allows for a modification of the central mass distribution through radiative cooling and feedback processes, resulting in larger concentrations (e.g. \citealt{mead10,fideli12}). Finally, it is worth mentioning the work by \cite{waizmann12,waizmann14}, who used the statistic of extreme values to show that very large Einstein radii are rare but not in conflict with $\Lambda$-CDM cosmology.\\
\indent By comparing observational results with numerical predictions, a better picture of the haloes' mass distribution has emerged, in particular for the sub-population of strong lenses and the 'problem' of their large concentrations. However, most of the studies have been focused so far on cluster-scale objects. Therefore, we propose here to extend the analysis of the mass-concentration relation of strong lenses towards lower mass scales. Based on a sample of objects detected and selected via their strong lensing signal, our study focuses on haloes of masses $\sim10^{13}-10^{14}\mathrm{M_{\odot}}$, with concentrations derived from a stacked weak-lensing analysis. Our main goal is to perform the first analysis of the strong-lensing bias from galaxy groups to massive clusters, and to compare the resulting mass-concentration relation with those derived from numerical simulations that mimic lensing-based analyses on sample of lensing-selected haloes.\\

This paper is organized as follows. In Section 2, we briefly present the data used in this work. We define in Section 3 the SARCS composite lenses, and introduce the specifics of a stacked weak-lensing analysis. The fitting results of the average shear profiles are given in Section 4, along with a discussion of the several simplifying hypothesis and sources of uncertainty introduced in our method. Section 5 is dedicated to the analysis of the groups' concentrations and the mass-concentration relation derived when combining our sample with massive strong-lensing galaxy clusters. Finally, we summarize our findings in Section 6. Throughout this paper, we use a standard $\Lambda$-CDM cosmology defined by $\Omega_\mathrm{M}=0.3, \Omega_\Lambda=0.7$, and a Hubble constant $H_0 = 70\,\mathrm{km/s/Mpc}$. Unless specified otherwise, masses are given in units of $\mathrm{M_{\odot}}$. Therefore, they should be re-scaled by our assumed $h=(H_0/100)=0.7$ before a comparison with results derived using a different $\Lambda$-CDM cosmology.

\section{The SARCS strong lensing galaxy groups}

\subsection{A sample of group-scale strong lenses}

To study the mass profile of strong-lensing galaxy groups, we use the Strong Lensing Legacy Survey sample (SL2S, \citealt{cabanac07}) and the recent compilation of its candidates with a group-scale gravitational arc (SARCS, \citealt{more12}). The sample was constructed with a semi-automated searching of elongated and curved features on the full Canada-France-Hawaii Telescope Legacy Survey (CFHTLS). The {\tt arcfinder} algorithm \citep{alard06,more12} was used to scan the 150 deg$^{2}$ of the CFHTLS optical images, leading to a number of $\sim1000$ candidates per square degree. After a visual inspection and a selection based on the quality ranking attributed to each potential lens, a sample of 127 systems was obtained. A photometric redshift was assigned to each candidate using the catalog of \cite{coupon09}, and different observing campaigns \citep{sl2s,thanjavur10,ruff11,munoz13} provided a spectroscopic redshift for several systems. More details about the CFHTLS data and the definition of the SARCS sample can be found in \cite{more12}.

\subsection{Most secure galaxy group candidates}

In \cite{foex13} (Paper I hereafter), we performed a weak-lensing and optical analysis of the SARCS sample of lens candidates, which provided two supplementary selection criteria to reduce the contamination of the sample by false detections and galaxy-scale lenses. With the fit of the systems' shear profile, we estimated their velocity dispersion, $\sigma_{v}$, via the Singular Isothermal Sphere mass model (SIS hereafter). We obtained a positive weak-lensing detection with $\sigma_{v}>0$ at the $1\sigma$ level for 89 objects. On the other hand, the study of the candidates' luminosity map using the galaxies populating the red sequence led to a total of 96 objects with an evident light over-density associated to the strong-lensing system. The combination of these two criteria resulted in a sample of 80 objects, which is the basis of the present analysis. These most secure lens candidates span large ranges in redshift ($z\in[0.15-1.2]$), mass ($\sigma_{v}\in[300-1100]$ km/s), and arc radius ($\mathrm{R_{A}}\in[2-20]''$). The average properties of this sample are given in the first row of Table 1, and we invite the reader to see Paper I for further details.

\section{Methodology}
\subsection{Advantages of a combining lenses}
As mentioned above, we performed in Paper I the weak-lensing analysis of each SARCS galaxy group candidates. However, the lensing signal-to-noise ratios (S/N hereafter) we measured were not high enough to derive reliable constraints on the mass of individual systems. To overcome this problem, we propose here to conduct a 'stack' analysis, i.e. combining several objects together to derive the properties of composite (average) galaxy groups. The main limitation in obtaining well constrained weak-lensing masses comes from the noise due to the galaxies' intrinsic ellipticity, whose dispersion is $\sim0.2-0.3$. Recovering a shear signal with intensities of $\sim0.1-0.01$ is a difficult task and requires averaging the shape of a large number $N$ of lensed galaxies. Since this noise scales as $1/\sqrt{N}$, by combining several lenses, one can artificially increase the source density and derive a shear signal with higher confidence levels. For instance, \cite{okabe13} stacked 50 galaxy clusters and obtained a shear profile with a total S/N of $\sim30$, compared to detection peaks of $\sim4$ in the two-dimensional mass map of individual objects (see also \citealt{okabe10a,oguri12,umetsu11b,umetsu14}).\\
\indent Stacking several lenses increases S/N ratios, but it is not the only improvement it provides (e.g. \citealt{oguri11}). When it comes to one-dimensional analyses, i.e. based on shear profiles, the assumption of circular symmetry can lead to biased estimates because of elliptical projected mass distributions (e.g. \citealt{corless08,feroz12}). By combining several lenses, the projected ellipticity of the individual objects gets averaged out, and a simple profile can provide a good description of the resulting average mass distribution. Furthermore, stacking several lenses reduces the impact of significant substructures in individual systems. In Paper I, we found 13 galaxy groups with complex light distributions, presenting two or more significant over-densities in their luminosity map. When combining together such systems, along with 'regular' ones, these substructures are naturally averaged, and their influence on the shear signal gets diluted.\\
\indent Finally, we can mention a last advantage of stacking/averaging several lenses. Weak-lensing deformations of source galaxies are produced by all the matter along the line of sight: large-scale structures (not correlated to the target lens) contribute to the shear signal, and can lead to biased mass estimates. Due to the linearity of the shear, the contribution of these structures simply adds to the signal produced by the lens. Because such contributions can be positive or negative, they get averaged out when combining several lenses (assuming an isotropic Universe), and only produce an additional statistical noise to the measured signal (e.g. \citealt{hoekstra01,hoekstra03}).

\subsection{Composite SARCS galaxy groups}
While the advantages of a stacking analysis are evident, one has to be careful with how to combine the lenses. A choice must be made between an increased S/N and a loss of information about the properties of the underlying population of lenses. As we have shown in Paper I, the SARCS sample is mainly made of group-scale lenses. However, it covers quite a large range in mass, up to galaxy clusters (arc radius up to 20'', $\sigma_{v}$ up to $\sim1000$ km/s). Therefore, we have chosen here to divide the sample in several stacks, which provides more data points to fit the mass-concentration relation (Section 5).\\
\indent To select which objects can be stacked together, while trying to reduce the scatter around the resulting composite lens, we use four selection criteria based on the individual properties of the groups. In Paper I, we derived for each object an estimate of the optical richness $N$ and luminosity $L$, using the bright galaxies populating the red sequence and located within a projected radius of 1 Mpc from the strong-lensing system. Given the scaling relations between these two quantities and the total mass of a galaxy group, we expect these two observables to provide a fairly good way of stacking objects according to their mass. Despite low S/N ratios of the shear signal, we also have an estimate of the SIS velocity dispersion $\sigma_{v}$, which we expect to be the most robust way to combine objects of similar mass. Finally, we have the direct observable of the arc radius $R_{A}$, values estimated in \cite{more12}. We have shown in \cite{verdugo14} that $R_{A}$ correlates with the groups' total mass, even though a large intrinsic scatter was found. This quantity is tightly related to the central mass distribution, thus, we expect it to be a better tracer of the dark matter halo concentration.\\
\indent After several tests, we decided to divide the sample of 80 objects into three stacks for each of the four selection criteria (richness, luminosity, velocity dispersion, and arc radius). In doing so, we obtained three uncorrelated points for the mass-concentration relation, while keeping a fairly high S/N of the stacked signal. We chose to put more objects in the low stacks (N1, L1, V1, and R1) because the lower mass lenses produce a more noisy shear signal. The limits of the middle and high stacks were chosen to have a similar number of objects, and to avoid a $1\sigma$ overlapping of the corresponding average selection criterion, i.e. the richness $\left<N\right>\pm1\sigma$ of the N stacks do not overlap, as for the $\left<L\right>$ of the L stacks, $\left<\sigma_{v}\right>$ of the V stacks, and $\left<R_{A}\right>$ of the R stacks. The general properties of the different stacks are given in Table 1 (using the individual properties derived in Paper I).

\begin{table*}
\centering 
\label{table:stacks}
\begin{threeparttable}
\caption{General properties of the different stacks.}
\begin{tabular}{l c c c c c c c c}
\hline\hline\noalign{\smallskip}
Stack ID & selection criterion & $\mathrm{N_{lens}}$ & $\left<z_{\mathrm{lens}}\right>$ & $\left<N\right>$ & $\left<L\right>\,(10^{12}\,\mathrm{L_{\odot}})$ & $\left<\sigma_{v}\right>\,(\mathrm{km/s})$ & $\left<R_{A}\right>\,(\mathrm{''})$\\
\noalign{\smallskip}\hline\noalign{\smallskip}
S0 & - & 80 & $0.55\pm0.19$ & $27\pm17$ & $1.97\pm1.22$ & $611\pm188$ & $4.3\pm3.1$\\
\hline
N1 & $5\leq N<20$ & 35 & $0.53\pm0.19$ & $12\pm3$ & $1.08\pm0.69$ & $531\pm163$ & $3.2\pm1.0$\\
N2 & $20\leq N<40$ & 22 & $0.60\pm0.21$ & $27\pm5$ & $2.03\pm0.79$ & $599\pm155$ & $3.6\pm2.0$\\
N3 & $40\leq N<75$ & 23 & $0.54\pm0.15$ & $50\pm9$ & $3.28\pm0.94$ & $746\pm178$ & $6.6\pm4.4$\\
\hline
L1 & $0.45\leq L<1.5$ & 37 & $0.51\pm0.16$ & $14\pm6$ & $0.92\pm0.31$ & $502\pm107$ & $3.6\pm1.5$\\
L2 & $1.5\leq L<3$ & 26 & $0.53\pm0.18$ & $33\pm11$ & $2.27\pm0.45$ & $629\pm161$ & $3.9\pm2.1$\\
L3 & $3\leq L<6$ & 17 & $0.68\pm0.18$ & $47\pm15$ & $3.81\pm0.70$ & $821\pm178$ & $6.5\pm5.1$\\
\hline
V1 & $300\leq \sigma_{v}<550$ & 35 & $0.55\pm0.18$ & $19\pm12$ & $1.38\pm0.85$ & $441\pm72$ & $3.5\pm1.4$\\
V2 & $550\leq \sigma_{v}<700$ & 23 & $0.46\pm0.14$ & $28\pm15$ & $1.81\pm0.98$ & $635\pm37$ & $4.1\pm2.4$\\
V3 & $700\leq \sigma_{v}<1100$ & 22 & $0.65\pm0.18$ & $39\pm18$ & $3.01\pm1.19$ & $857\pm110$ & $5.7\pm4.6$\\
\hline
R1 & $2\leq R_{A}<3.5$ & 35 & $0.56\pm0.21$ & $21\pm14$ & $1.62\pm1.03$ & $565\pm162$ & $2.6\pm0.5$\\
R2 & $3.5\leq R_{A}<5.5$ & 29 & $0.55\pm0.18$ & $27\pm14$ & $2.00\pm1.01$ & $618\pm207$ & $4.2\pm0.6$\\
R3 & $5.5\leq R_{A}<20$ & 14 & $0.52\pm0.14$ & $44\pm18$ & $2.66\pm1.53$ & $704\pm181$ & $9.4\pm4.0$\\
\noalign{\smallskip}\hline
\end{tabular}
    \begin{tablenotes}
      \small
      \item Columns are (1) name of the stack; (2) selection criterion; (3) number of lenses in the stack; (4) average redshift; (5)-(6) average richness and optical luminosity within 1 Mpc (red sequence galaxies, $M_{i'}<-21$); (7) average weak-lensing SIS velocity dispersion; (8) average arc radius. Means were derived using the groups' individual properties given in Paper I. Note: 2 objects with no estimate of $\mathrm{R_{A}}$ were not included in the R-stacks.
    \end{tablenotes}
  \end{threeparttable}
\end{table*}

\subsection{Weak-lensing stacked analysis}
Our weak-lensing pipeline is described in Paper I, and we review here some details of a stacked analysis.\\
\indent Using the second derivative of the projected gravitational potential to express the shear and convergence, one can show that for a lens with a circular-symmetric projected mass distribution, the two weak-lensing deformations are simply related through \citep{miralda91}:
\begin{equation}
\gamma_{t}(r)=\overline{\kappa}(<r)-\overline{\kappa}(r),
\end{equation}
where $\overline{\kappa}(<r)$ and $\overline{\kappa}(r)$ are the convergence averaged over the disk and circle of radius $r$, respectively. Since the convergence $\kappa$ is equal to the surface mass density $\Sigma(r)$ normalized by a critical density $\Sigma_{\mathrm{crit}}$, we can rewrite the previous equation as:
\begin{equation}
\label{eq:dsig}
\overline{\Sigma}(<r)-\overline{\Sigma}(r)=\Sigma_{\mathrm{crit}}\times\gamma_{t}(r),
\end{equation}
The critical density reads:
\begin{equation}
\Sigma_{\mathrm{crit}}=\frac{c^{2}}{4\pi G}\frac{1}{\beta D_{OL}},
\end{equation}
where the factor $\beta=D_{LS}/D_{OS}$ captures the geometrical configuration of the lensing optical bench. From Equation \ref{eq:dsig}, we see that the shear produced by a lens equals its mass density contrast $\Delta\Sigma(r)\equiv\overline{\Sigma}(<r)-\overline{\Sigma}(r)$ after a rescaling by the critical density. In other words, one can combine (average, stack) the shear signal produced by lenses at different redshifts to recover the mass of the corresponding composite lens.\\
\indent The density contrast of a circular-symmetric lens $j$ can be locally estimated by the tangential shear $\gamma_{t,ij}$ it produces on a galaxy source $i$ located at the concentric radius $r_{ij}$:
\begin{equation}
\Delta\tilde{\Sigma}_{j}(r_{ij})=\Sigma_{\mathrm{crit},ij}\times\tilde{\gamma}_{t,ij}.
\end{equation}
To derive the shear, we employ the estimator $\tilde{\gamma}_{t}(r)=\left<e_{\|}\right>$, i.e. the average tangential ellipticity component of the background galaxies located at a radius $r\pm\delta r$. Since we do not have an estimated redshift for each galaxy, we use the same critical density $\Sigma_{\mathrm{crit},j}$ for all the source galaxies $i$ of a given lens $j$. This critical density is calculated with the average geometrical factor $\left<\beta(z)\right>$, whose values are given in Paper I for each SARCS lens.\\
\indent By combining the signal of several lenses, we increase the number of available sources for the resulting composite object and reduce the noise due to the galaxies' intrinsic ellipticity. The corresponding average mass density contrast reads:
\begin{equation}
\label{eq:sum}
\left<\Delta\tilde{\Sigma}(r)\right>=\frac{\sum_{j=1}^{N_{\mathrm{Lens}}}\sum_{i=1}^{N_{\mathrm{Sources}}}\omega_{ij}\times e_{\|,ij}\times\Sigma_{\mathrm{crit},j}}{\sum_{j=1}^{N_{\mathrm{Lens}}}\sum_{i=1}^{N_{\mathrm{Sources}}}\omega_{ij}},
\end{equation}
where $N_{sources}$ is the number of source galaxies within the annulus of projected physical radius $r\pm\delta r$ around the centre of the $j$th lens.\\
\indent In order to reduce the impact of galaxies with a noisy estimate of their shape parameters, the tangential component of the ellipticity is weighted according to the inverse variance of its measurement:
\begin{equation}
\omega_{ij}=\frac{1}{(\Sigma_{\mathrm{crit},j}\times\sigma_{e_{\|,ij}})^{2}}.
\end{equation}
Therefore, the statistical uncertainty associated to our estimator $\Delta\tilde{\Sigma}(r)$ can be expressed as:
\begin{equation}
\label{eq:disp}
\sigma_{\Delta\tilde{\Sigma}}^{2}=\frac{\sum_{i}\omega_{i}^{2}\times\sigma_{\tilde{\gamma}_{t,i}}^{2}\times\Sigma_{\mathrm{crit},i}^2}{(\sum_{i}\omega_{i})^{2}},
\end{equation}
where the sum runs over all the stacked background galaxies in the radial bin $r$. The uncertainty on the tangential shear $\sigma_{\tilde{\gamma}_{t,i}}$ is given by the quadratic sum of the errors on the shape measurement $\sigma_{e_{\|,ij}}$ and the noise due to the galaxies' intrinsic ellipticity, derived assuming an RMS of 0.25 per component.\\
\indent To quantify the detection level of the signal for a given stack, we define the total S/N ratio as follows:
\begin{equation}
\label{eq:sn}
\left(\frac{\mathrm{S}}{\mathrm{N}}\right)^2=\sum_i\left(\frac{\left<\Delta\tilde{\Sigma}(r_i)\right>^2}{\sigma_{\Delta\tilde{\Sigma}}^{2}}\right),
\end{equation}
where the sum runs over the bins in radius used to fit the profile. Here we only consider the statistical uncertainty defined above as source of noise, an approximation justified in Section 4.2. The total S/N ratios of the profiles are given Table 2; stacking the 80 SARCS galaxy groups leads to $S/N=14.3$ over the range 50-3000 kpc.\\
\indent With Equation \ref{eq:sum}, we have an estimator of the mass density contrast for a stack of lenses. However, the signal that is actually measured when averaging the shape of lensed galaxies is the reduced shear $g=\gamma/(1-\kappa)$. Therefore, one cannot simply fit the stacked signal by the analytical expression $\Delta{\Sigma}(r)$ of a given mass model. Since we want to take full advantage of the stacking procedure to obtain constraints in the central regions of the lenses where the weak-lensing approximation $g\approx\gamma$ no longer holds, we need to evaluate what our estimator actually measures. It can be shown that, indeed, the estimator $\Delta\tilde{\Sigma}(r)$ has a second-order contribution \citep{mandelbaum06,johnston07}: 
\begin{equation}
\label{eq:fit}
\Delta\tilde{\Sigma}(r)=\Delta\Sigma(r)+\Delta\Sigma(r)\times\Sigma(r)\times L_{Z},
\end{equation}
with
\begin{equation}
\label{eq:lz}
L_{Z}=\frac{\left<\Sigma_{\mathrm{crit}}^{-3}\right>}{\left<\Sigma_{\mathrm{crit}}^{-2}\right>}.
\end{equation}

Neglecting variations in the density of source galaxies between the different radial bins (e.g. \citealt{johnston07,leauthaud10}), we estimate for each composite lens an average factor $L_{Z}$ over the range 0.1-2 Mpc from the lens centre. Following the methodology used in Paper I to derive the average geometrical factor $\left<\beta\right>$, we estimate the ratio $\left<\beta^{3}\right>/\left<\beta^{2}\right>$ for each individual group. We then calculate $L_{Z}$ for a given composite lens as:

\begin{equation}
\label{eq:lzsum}
L_{Z}=\frac{4\pi G}{c^{2}}\frac{\sum_{j=1}^{N_{\mathrm{Lens}}}N_{\mathrm{gal},j}\left<\beta^{3}\right>_{j}D_{OL,j}^{3}}{\sum_{j=1}^{N_{\mathrm{Lens}}}N_{\mathrm{gal},j}\left<\beta^{2}\right>_{j}D_{OL,j}^{2}},
\end{equation}
with $N_{\mathrm{gal}, j}$ the number of source galaxies within 0.1-2 Mpc from the centre of the $j$th group. The value of $L_{Z}$ for each stack is given in Table 2.

\section{Mass profile of composite galaxy groups}

\subsection{Modeling the data}

Most of the stacked weak-lensing analyses make use of the so-called 'halo model' (e.g. \citealt{mandelbaum05b,mandelbaum06,johnston07,mandelbaum08b,mandelbaum10,leauthaud10,oguri11,covone14,umetsu14}). With this approach, the mass density contrast is modeled as the sum of three components: the stellar mass contained in the central galaxy, the group- or cluster-scale main halo (the 'one-halo term'), and a contribution from other groups and clusters surrounding the target (the 'two-halo term'). While the first term only produces a significant contribution on very small scales (typically below 50 kpc), the two-halo term only has a dominant contribution well beyond the Virial radius of the main halo (typically several Mpc, e.g. \citealt{oguri11}). On intermediate scales, the signal is largely dominated by the contribution of the one-halo term. In the present work, we restrict our analysis to the one-halo term only; thus, the expressions of $\Delta\Sigma(r)$ and $\Sigma(r)$ (in Equation \ref{eq:fit}) do not include a stellar contribution nor the large-scale two-halo term. We discuss in the next subsection the validity of this approximation.\\
\indent To fit the observed density-contrast profiles of the SARCS galaxy groups, we employ three mass models. First, we use the SIS, which is fully characterized by its velocity dispersion $\sigma_{v}$, and has a mass-density profile with a constant logarithmic slope of -2. While this model has proven to give a good description for the mass distribution of individual galaxies, it is not expected to reproduce accurately more massive dark matter haloes, which are expected to present a steeper density profile at large scales. However, the SIS velocity dispersion $\sigma_{v}$ can be easily compared to results from a dynamical analysis (e.g. \citealt{munoz13}) or strong-lensing models providing an estimate of the Einstein radius (e.g. \citealt{verdugo14}). The shear produced by a SIS scales as $\gamma(r)\propto\sigma_{v}^{2}r^{-1}$.\\
\indent The second model we use is a mass distribution characterized by a power law density profile with a constant but free slope (PLAW). We express its surface mass density as $\Sigma(r)=\Sigma_{0}r^{\alpha}$, with $\Sigma(1\,\mathrm{Mpc})=\Sigma_{0}$. A slope $\alpha=-1$ corresponds to the SIS model. The density contrast of the PLAW model reads $\Delta\Sigma(r)=(-\alpha/(2+\alpha))\Sigma_{0}r^{\alpha}$.\\
\indent Finally, we use the NFW mass profile, derived from extensive dark matter numerical simulations. This model is supposed to reproduce the mass distribution of dark matter haloes over a wide range in mass, from galaxy to cluster scales \citep{navarro95,navarro96,navarro97}. Its density reads:
\begin{equation}
\label{eq:rhonfw}
\rho(r)=\frac{\rho_{0}}{(r/r_{s})(1+r/r_{s})^{2}}.
\end{equation}
The scale radius $r_{s}$ marks the transition between the two asymptotic behaviors, $\rho\propto r^{-1}$ in the central part, and a steeper profile $\rho\propto r^{-3}$ in the outskirts. Thus, the NFW model provides more freedom to characterize the mass profile, contrary to the two other models whose slopes are constant at all radii. The normalization of the NFW profile $\rho_{0}$ is related to the 3D mass via:
\begin{equation}
\label{eq:nfw}
M(<r)=4\pi\rho_{0}^{3}r_s^3\left[\ln(1+r/r_{s})-(r/r_{s})/(1+r/r_{s})\right].
\end{equation}
\indent In the model of the gravitational collapse of a spherical over-density, one can show that a virialized object reaches a density contrast $\Delta_{vir}\approx180$ with respect to the mean density $\overline{\rho}_{m}$ of an Einstein-De Sitter Universe \citep{gunn72}; \cite{bryan98} derived an accurate approximation for $\Delta_{vir}(z)$ in a $\Lambda$CDM Universe. This simple prescription gives a natural parametrization of the NFW profile: the radius within which the dark matter halo's averaged density equals $\Delta_{vir}(z)\overline{\rho}_{m}(z)$ defines the Virial mass,
\begin{equation}
\label{eq:mvir}
M_{vir}\equiv M(<R_{vir})=\frac{4\pi}{3}R_{vir}^3\Delta_{vir}(z)\overline{\rho}_{m}(z).
\end{equation}
This mass can be combined with Equation \ref{eq:nfw} to simply express the normalization of the NFW profile as $\rho_{0}=\delta_{c}\overline{\rho}_{m}(z)$, where the characteristic over-density $\delta_{c}$ equals:
\begin{equation}
\label{eq:dc}
\delta_{c}=\frac{\Delta_{vir}(z)}{3}\frac{c_{vir}^{3}}{\ln{(1+c_{vir})}-c_{vir}/(1+c_{vir})},
\end{equation}
with the concentration parameter $c_{vir}=R_{vir}/r_{s}$. It is thus possible to express the NFW profile in terms of $(M_{vir},c_{vir})$ rather than its normalization $\rho_{0}$ and scale radius $r_{s}$. Different parametrizations can be found in the literature. For instance, instead of using the Universe's mean density $\overline{\rho}_{m}(z)$, one can use the critical density, $\rho_{c}(z)=\overline{\rho}_{m}(z)\Omega_{m}(z)^{-1}$, which can be justified as follows: in a $\Lambda$CDM Universe, an over-density will collapse if it behaves as a mini-closed Universe, i.e. having an average density larger than the critical density $\rho_{c}$. It is also common to use a fixed density contrast for every redshift, often set to $\Delta=200$, a choice that can be motivated because of the cosmology dependence of $\Delta_{vir}(z)$, and the assumption that galaxy groups and clusters are not necessarily virialized at the time we observe them. Even though these different parametrizations complicate the comparison between different studies, it is easy to convert results from one definition to another by combining Eq. \ref{eq:mvir}, Eq. \ref{eq:dc}, and keeping constant the absolute normalization $\rho_{0}$ (e.g. \citealt{hukravtsov03}). The NFW mass model has an analytical expression for the shear (e.g. \citealt{bartelmann96}), and in the rest of the paper, we will use the $(M_{200},c_{200})$ parametrization, with the density contrast expressed with respect to the critical density, i.e. $\rho_{0}=\delta_{c}\rho_{c}(z)$.\\
\indent Even though the NFW model provides a fairly good description of the galaxy groups and clusters' mass distribution, several improvements have been proposed, including triaxiality, core of constant density, or varying inner logarithmic slope (e.g. generalized NFW, Einasto profile). However, estimating the extra free parameters of such models requires observational data in the central part of the halo, a region not accessible with the present weak-lensing observations (see e.g. \citealt{newman09,newman13} for a combination of lensing and stellar kinematics to probe the mass profile down to $\sim$ kpc scales). Therefore, we limit our analysis to the classical NFW profile.

\subsection{Error budget - Validation of the method}

Any weak-lensing study comes with several sources of both statistical and systematic error (e.g. \citealt{mandelbaum05a}). We review in the following the dominating ones and we describe how we adapted our methodology accordingly.\\

{\it Shear calibration}\\
In \cite{foex12}, we ran our lensing pipeline on the STEP1 simulations \citep{heymans06}, and we derived a calibration bias of $-0.1\pm0.02$. Therefore, prior to the fit of the shear profiles, we corrected our measured signal by a boost factor of $10\%$. To further test whether our lensing measurements suffer from residual systematics, we computed shear profiles using the radial component of the galaxies' ellipticity. Figure \ref{fig:stacks} shows that this signal, expected to be null, is indeed statistically consistent with 0 (within $3\sigma$ at most) over the range in radius used to fit the profiles.\\

{\it Redshift distribution of the sources}\\
To translate the geometrical weak-lensing signal into the mass of the deflector, one has to evaluate the average geometrical factor $\left<\beta(z_{l},z_{s})\right>$, which depends on the redshift distribution of the lensed galaxies. Our approach, described in Paper I (see also \citealt{sl2s,foex12}), makes use of photometric redshifts that were carefully calibrated with spectroscopic observations \citep{ienna06}. Because they were derived with the same CFHTLS observations used for this work, it is straightforward to apply the selection criteria (magnitude and color) to these catalogs, and derive the redshift distribution of the lensing sources. We verified that differences in the value of $\left<\beta\right>$ are typically of the percent level when using the redshift distribution of the different CFHTLS fields. It is much lower than the statistical noise due to the dispersion of the galaxies' intrinsic ellipticity, so this source of uncertainty can be neglected. In Paper I, we also investigated the influence of uncertainties in the lens redshift $z_{l}$, and we found that an error of $|z_{l}-z_{l}^{true}|=0.1$ propagates to a $20\%-30\%$ error on the mass. The comparison between spectroscopic and photometric redshifts has shown that the overall agreement is better than this 0.1 uncertainty (Fig. 2 of Paper I); therefore, we can assume that the optical benches of the stacks are well enough constrained to not generate significant errors in the mass estimates.\\

{\it Centre offset}\\
When using shear profiles, the position of the mass centre has to be carefully chosen. Indeed, a wrong centre acts as a smoothing of the shear signal, which leads to biased-low mass and concentration estimates (e.g. \citealt{johnston07,george12,covone14}). A dark matter halo is by definition dark and not visible, thus, locating its mass centre can be a challenging task. The weak-lensing signal can be used itself to constrain the centre position, but low spatial resolution, sparse constraints and noisy data can lead to biased estimates, in particular with ground-based observations (e.g. \citealt{dietrich11}). Baryonic tracers are usually employed to locate the mass centre, in particular with the position of the brightest central galaxy. However, this method suffers from two assumptions: that the brightest galaxy is correctly identified, and that it lies at the actual centre of mass (see discussions in \citealt{johnston07} and \citealt{mandelbaum08b} for more details). \cite{george12} compared different approaches to estimate the mass centre, for instance, by using the centroid of the galaxy population rather than the brightest one, or including stellar mass and X-ray emission information: while the centre of systems with a clear central galaxy is fairly well traced by its position (offsets smaller than 75 kpc), in most cases the best tracer is obtained by the position of the brightest/most massive galaxy close to the peak of the X-ray emission. However, such a method requires X-ray observations, so we cannot use it here. On the other hand, for systems going through a major merging event, the X-ray emission peak can be dislocated from the actual mass centre, as observed in the so-called 'Bullet' clusters \citep{markevitch04,bradac08b,merten11,dawson12,dahle13}. Recently, we found such a bullet-like object in the SARCS sample, the lowest mass system observed to date with a displacement between the X-ray emission peak and the mass centre \citep{gastaldello14}. A parallel study based on numerical simulations showed that such low-mass systems are more frequent than bullet configurations in massive galaxy clusters \citep{fernandez14}. Thus we can expect that, although efficient in most cases, the actual mass centre cannot be accurately traced by the X-ray emission for a non negligible fraction of galaxy groups.\\
\indent Since gravitational lensing does not rely on the baryonic content of a dark matter halo, the strong-lensing system can be expected to be an accurate tracer of the mass centre, even in such dynamically perturbed galaxy groups (see Section 3.2 and Appendix of Paper I for further details). Therefore, we use the centre of the SARCS gravitational arcs to define the centre of the profiles, and we do not include in our fitting procedure a contribution to the signal by objects with a wrongly-identified mass centre (see \citealt{johnston07} for a method that accounts for centre offsets in a stacked weak-lensing analysis). Furthermore, in the case of strong-lensing events produced by a sub-halo of a more massive component, the weak-lensing signal of the larger-scale object will mainly contribute to the two-halo term of the stacked shear profiles. As shown below, we can neglect this term in our analysis, thus, our choice for the mass centre provides a globally consistent approach.\\

 \begin{figure}[!t]
\center
\includegraphics[width=\hsize, angle=0]{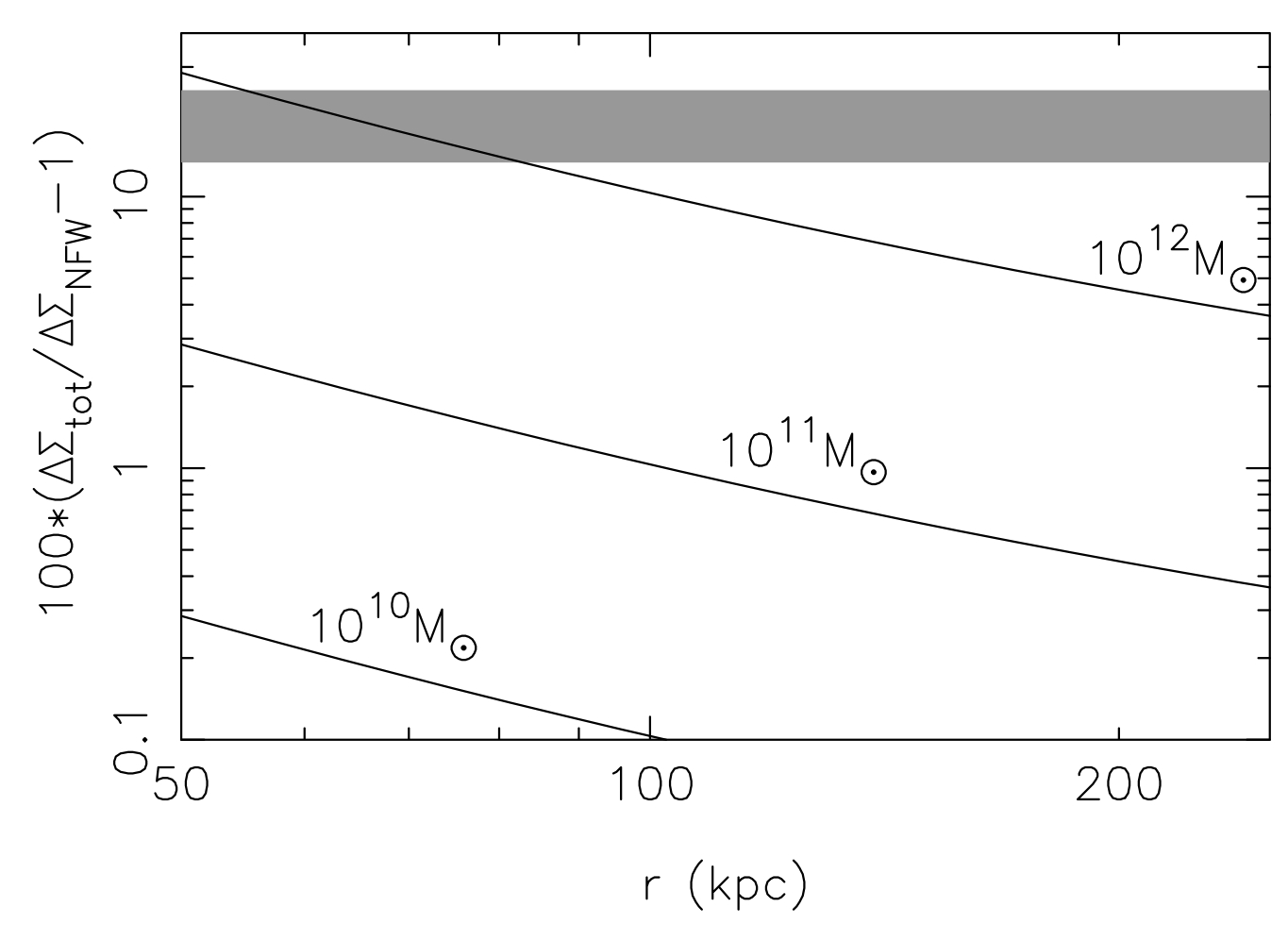}
\caption{Relative contribution of a point mass component to the total measured signal, computed with Equation \ref{eq:fit}, and using a NFW halo described by the best-fit parameters of the stack S0. The three curves show the contribution produced by a mass of 0.01, 0.1, and $1\times10^{12}\mathrm{M_{\odot}}$. The grey area represents the typical relative uncertainty of the measured signal in the region 50-300 kpc.}
\label{fig:PM} 
\end{figure}

{\it Mass modeling}\\
The three mass models we use do not have any explicit contribution for the baryonic content of galaxy groups. A more complete description of their mass distribution should in fact account for the presence of the gas, and the stellar population within the galaxy members. For the former, one can argue that its spatial distribution follows that of the dark matter (for relaxed objects at least), thus, its presence does not significantly modify the shape of the total mass density profile. For the stellar mass contribution, one can assume as well that the distribution of satellite galaxies follows that of the dark matter (e.g. \citealt{leauthaud11}). Therefore, to first order, we can consider that the total mass distribution of our composite lenses is well represented by a single NFW component.\\
\indent On the other hand, a massive central galaxy introduces a contribution to the lensing signal that needs to be accounted for because it can bias the results of the NFW fitting (e.g. \citealt{johnston07,leauthaud10}). The presence of baryons in the core of a dark matter halo can further modify its density profile by adiabatic contractions (e.g. \citealt{gnedin04}). However, we can expect that for the present data quality, our fitting method does not provide statistically significant differences in the best fit parameters when accounting or not for an additional central mass component (see also \citealt{okabe13,umetsu14}). Because we fit the shear profiles above 50 kpc, where the mass density of the central galaxy is negligible, we can model this component as a simple point mass rather than a more complex mass distribution, e.g. a Sersic profile. Figure \ref{fig:PM} shows the relative contribution of such a point mass with $\Delta\Sigma(r)=M_{0}/\pi r^{2}$ to the signal produced by a NFW halo characterized by the S0 best-fit parameters. Above 100 kpc, a point mass of $M_{0}=10^{12}M_{\odot}$ contributes less than $10\%$ to the total signal. This value is lower than the typical uncertainty of our weak-lensing measurements over the range 50-300 kpc, which justifies not including this component in the model. In Section 4.4, we investigate anyway the contribution of a point mass, when combining the stacked weak-lensing signal with strong-lensing constraints. Even though the central galaxy can bias the mass estimate of the dark matter halo itself, numerical (e.g. \citealt{laporte14}) and observational (e.g. \citealt{newman13}) studies have shown that its presence leads to a the total mass profile that follows the NFW model. This consideration further justifies the use of a simple single-component mass distribution.\\
\indent The other main approximation in our mass modeling consists in neglecting the two-halo term. For instance, \cite{umetsu14} have shown that this component produces a signal slowly decreasing with increasing radius, therefore not affecting significantly the shape of the weak-lensing profile. Moreover, since we limit the profiles to 3 Mpc, most of the constraints in the fit come from radii in the range 0.1-1 Mpc (up to $\sim\,\mathrm{R_{v}}$), a region where the two-halo term has a contribution $\sim$ 1-2 orders of magnitude smaller than that of the one-halo term (e.g. Figure 8 of \citealt{johnston07}). Since we have a $\sim20\%$ uncertainty on the measured signal in the same region, we can safely neglect this contribution in our mass modeling.\\
\indent Finally, it is worth mentioning that we only use spherical mass models, although dark matter haloes are known to be triaxial (see e.g. the review of \citealt{limousin12}). We explore the effects of this approximation on the mass-concentration relation in Section 5.2, and we provide a simple way to estimate the elongation along the line of sight of prolate haloes in Section 5.3.\\

{\it Large-scale structures}\\
As briefly discussed in Section 3.1, the signal produced by uncorrelated large-scale structures can bias the lensing mass of a single lens, but their extra deformations are averaged out when stacking several objects. However, they introduce a statistical noise in the lensing signal. As shown in \cite{oguri11} for stacked high-redshift galaxy clusters (Figure 8), the dominant source of noise up to $\sim 15$ arcmin is produced by the intrinsic ellipticity of lensed galaxies. At the average redshift of our stacks $z\sim0.5$, the limit in radius where we fit the profiles is 3 Mpc $\sim8$ arcmin. Therefore, we can neglect this source of noise in our calculations, as done in similar studies of stacked weak-lensing analysis of galaxy groups (e.g. \citealt{leauthaud10}).\\

\begin{figure}
\center
\includegraphics[width=\hsize, angle=0]{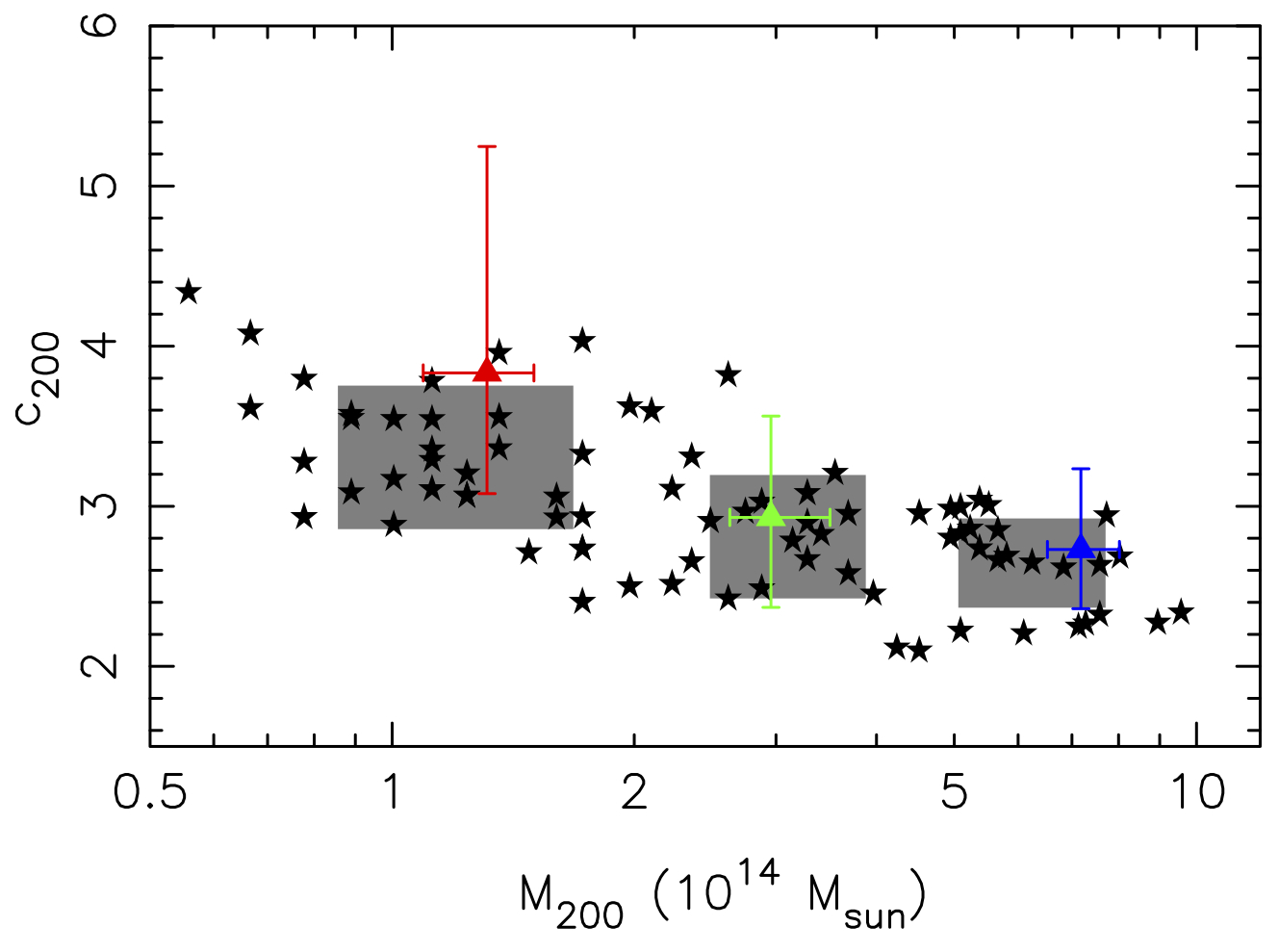}
\caption{Results of the stacked analysis on simulated catalogs of lensed galaxies that mimic our observations for the full sample of 80 strong lenses. Colored triangles show the best-fit mass and concentration for the three composite lenses corresponding to stacks in richness. The grey-shaded area covers the intrinsic dispersion of the simulated lenses (black stars) around their average mass and concentration. The overlap of these average values with the results of the stacked analysis indicate the absence of any strong systematic bias.}
\label{fig:simu} 
\end{figure}

{\it Stacking procedure}\\
The profiles are constructed with logarithmically-spaced annuli. We start to fit them at 50 kpc from the centre, thus limiting the influence of mis-centering and reducing the contribution of the central baryonic mass component. The outer limit of the profiles is set to 2 Mpc for the low stacks, and 3 Mpc for the middle and high ones, resulting in eight and nine bins, respectively. We verified that slightly changing the inner and outer limits of the profiles does not give statistically different best-fit parameters.\\
\indent To further test our stacking procedure, we did several statistical simulations, i.e. simulating catalogs of lensed galaxies, analyzing them with our method, and comparing the results with the expected values. The catalogs were generated with the Lenstool code \citep{kneib96,jullo07} as follows. Based on our list of 80 groups with measured richness N(r$<$1 Mpc), we assumed they were modeled with NFW density profiles, and their NFW parameters were scaled with the following relations between the richness and the mass \citep{mandelbaum08}
\begin{equation}
\label{eq:scale1}
M_{200}  = M_0(N/N_0)^A
\end{equation}
with $M_0=1.56\times 10^{14}\ M_\odot$, $A=1.15$ and $N_0=20$, and between the mass and the concentration 
\begin{equation}
\label{eq:scale2}
c_{200} = c_0(M_{200}/M_0)^{-B} (1+z)^{-0.45}
\end{equation}
with $B=0.13$ and $c_0=4.6$.\\
For each group, we simulated a catalog of about 3000 sources at redshift $z=1.171$, with an intrinsic Gaussian shape noise $e_{int} = 0.25$, and over an area of 15x15 arcmin$^{2}$. The results we obtained by stacking the groups according to their richness are presented in Figure \ref{fig:simu}, with masses and concentrations that are fully consistent with the average value of the simulated groups. Therefore, we can conclude that our methodology does not suffer any strong systematic bias.

\subsection{Fitting results of the observed $\Delta\Sigma(r)$ profiles}

 \begin{figure}
\center
\includegraphics[width=\hsize, angle=0]{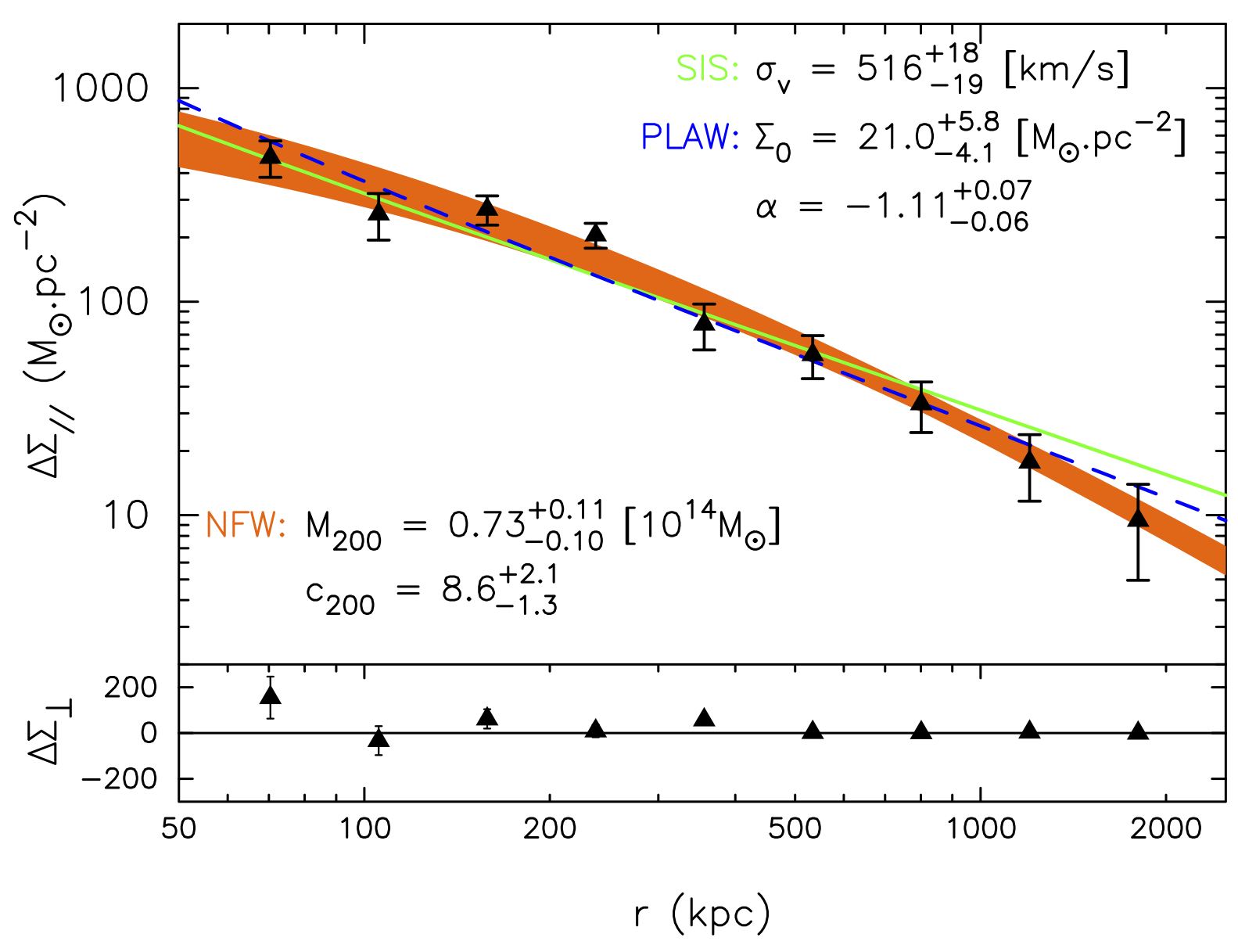}
\caption{Average density contrast $\Delta\Sigma(r)$ for the stack S0 (Equation \ref{eq:sum}). The lower panel shows the profile obtained using the radial component of the lenses galaxies, and should be equals to zero. The upper panel is the profile using the tangential component. Over-plotted are the best-fit results (right side of Equation \ref{eq:fit}) for the three mass models, SIS (green-solid line), PLAW (dashed-blue line) and NFW (orange-shaded area, encompassing the combined $1\sigma$ uncertainty on $M_{200}$ and $c_{200}$). }
\label{fig:stacks} 
\end{figure}

To derive the set of parameters $\pmb{\theta}$ that describes best the data, we perform a standard $\chi^{2}$ minimization:
\begin{equation}
\chi^{2}=\sum_{i}^{N}\frac{\left(\left<\Delta\tilde{\Sigma}(r_{i})\right>-\Delta\tilde{\Sigma}(r_{i},\pmb{\theta})\right)^2}{\sigma_{i}^{2}},
\end{equation}
where the sum runs over the N radial bins of the profile. The model prediction $\Delta\tilde{\Sigma}(r,\pmb{\theta})$ is derived from Equation \ref{eq:fit} using the corresponding analytical expressions for $\Delta\Sigma(r)$ and $\Sigma(r)$. The statistical uncertainties $\sigma_{i}$ (Equation \ref{eq:disp}) are propagated on the best-fit parameters with a Monte Carlo approach: we generate 10,000 new profiles, whose points are randomly drawn from the normal distributions $\mathcal{N}\left(\left<\Delta\tilde{\Sigma}(r_{i})\right>,\sigma_{i}^{2}\right)$. The new profiles are fitted with the previous equation, providing at the end an estimate of the probability distribution of the model's free parameters. The mode of the (marginalized) distribution gives the value of the best-fit parameter, and the associated $1\sigma$-confidence interval is given by the range encompassing 68\% of the drawings on each side. The fitting results for the SIS, PLAW, and NFW mass models are given in Table 2. Figure \ref{fig:stacks} shows the results obtained for the stack S0, and Figure \ref{fig:concentration} gathers the NFW masses and concentrations for the all the stacks.\\
\indent For every stack, the three mass models provide good fits, with reduced $\chi^{2}$ not exceeding $\sim2$ (except for the stack L1). For the stack S0, which has the largest difference in the fit quality between the SIS and NFW models, we obtain a likelihood ratio $\Delta\chi^{2}\equiv\chi^{2}_{SIS}-\chi^{2}_{NFW}=9$: the SIS model is only disfavored at the $3\sigma$ level. Moreover, we see that the slope of the PLAW model is fairly consistent with that of the SIS model (i.e. $\alpha=-1$), results suggesting that group-scale objects are well described by a SIS mass distribution in the range probed by our weak-lensing data. As shown by the study of arc statistics \citep{oguri06,more12}, strong-lensing galaxy groups are, indeed, expected to fall in between NFW massive galaxy clusters and SIS single galaxies. Previous stacked analyses of galaxy clusters resulted in SIS fits rejected with a higher significance: $11\sigma$ level in \cite{okabe10a}  (weak lensing, nine clusters, $M_{vir}=9.7\pm0.8\times10^{14}h^{-1}M_{\odot}$), $8\sigma$ in \cite{umetsu11b} (weak+strong lensing, four clusters, $M_{vir}=15.4\pm1\times10^{14}h^{-1}M_{\odot}$), $4\sigma$ in \cite{umetsu14} (weak lensing + magnification, 20 clusters, $M_{200}=13.4\pm1\times10^{14}M_{\odot}$). \cite{okabe10a} found a better agreement between the SIS and NFW fits when selecting objects with a lower mass, with a SIS fit disfavored at the $6\sigma$ level (10 clusters, $M_{vir}=4.8\pm0.4\times10^{14}h^{-1}M_{\odot}$), a value closer to our findings. Therefore, we can explain the relatively good agreement between the SIS and NFW fits that we obtain for the SARCS sample by the lower mass scales it covers.\\
\indent Even though we are not using an aperture scaling with mass or redshift to estimate the richnesses and luminosities, we observe the usual optical scaling relations: the mass parameters ($\sigma_{v}$, $\Sigma_{0}$, and $M_{200}$) increase for larger richnesses (N stacks) and luminosities (L stacks). As expected, the V stacks give the larger range in $M_{200}$: although very noisy, the individual velocity dispersions derived in Paper I provide the best way to stack objects with a similar total mass. The sample covers one order of magnitude in mass, from $M_{200}=0.21\pm0.07\times10^{14}M_{\odot}$ for the stack V1 to $M_{200}=2.38\pm0.5\times10^{14}M_{\odot}$ for the stack V3. On the other hand, the arc radius appears to be a poor tracer of the total mass, with mass parameters roughly constant for the three R stacks. This result is not surprising because strong-lensing features are only related to the central part of the lens, whose projected mass density does not necessarily scale with the total mass of the dark matter halo (see Section 5.2, and \citealt{verdugo14}). 

\begin{table*}
\centering 
\label{table:fits}
\begin{threeparttable}
\caption{Fitting results of the density-contrast profiles $\Delta\tilde{\Sigma}(r)$ (Equation \ref{eq:fit}) using the SIS, NFW, and PLAW mass models.}
\begin{tabular}{l c c c c c c c c c c c c c}
\hline\hline\noalign{\smallskip}
  & & & & \multicolumn{2}{c}{SIS} & & \multicolumn{3}{c}{PLAW} & & \multicolumn{3}{c}{NFW}\\
  \cline{5-6}\cline{8-10}\cline{12-14}
 Stack ID & $z_{\mathrm{stack}}$ & $L_{Z}$ & (S/N) & $\chi^{2}$/dof & $\left<\sigma_{v}\right>$ & & $\chi^{2}$/dof & $\left<\alpha\right>$ & $\left<\Sigma_{0}\right>$ & & $\chi^{2}$/dof & $\left<M_{200}\right>$ & $\left<c_{200}\right>$\\
   &  & ($10^{-4}\mathrm{pc^{2}.M_{\odot}^{-1}}$) & & & (km/s) &  &  & & $(\mathrm{M_{\odot}.pc^{-2}})$ &  & & $(10^{14}\,\mathrm{M_{\odot}})$ &\\   
\noalign{\smallskip}\hline\noalign{\smallskip}
S0 & 0.52 & 3.87 & 14.3 & 15.8/8 & $516_{-19}^{+18}$ & & 13.1/7 & $-1.11_{-0.06}^{+0.07}$ & $21.0_{-4.1}^{+5.8}$ & & 6.8/7 & $0.73_{-0.10}^{+0.11}$ & $8.6_{-1.3}^{+2.1}$\\[3pt]
\hline\noalign{\smallskip}
N1 & 0.47 & 3.80 & 6.8 & 11.8/7 & $410_{-36}^{+33}$ & & 11.2/6 & $-1.16_{-0.13}^{+0.13}$ & $9.8_{-3.5}^{+7.2}$ & & 8.2/6 & $0.36_{-0.08}^{+0.12}$ & $7.9_{-1.9}^{+4.6}$\\[3pt]
N2 & 0.59 & 3.87 & 7.7 & 10.2/8 & $554_{-42}^{+33}$ & & 10.2/7 & $-1.09_{-0.11}^{+0.14}$ & $22.0_{-6.5}^{+16.5}$ & & 10.8/7 & $0.72_{-0.16}^{+0.31}$ & $7.9_{-2.5}^{+4.8}$\\[3pt]
N3 & 0.53 & 3.98 & 12.6 & 3.8/8 & $672_{-30}^{+24}$ & & 4.2/7 & $-0.96_{-0.07}^{+0.07}$ & $54.9_{-12.0}^{+15.2}$ & & 4.0/7 & $1.64_{-0.27}^{+0.34}$ & $7.2_{-1.3}^{+2.0}$\\[3pt]
\hline\noalign{\smallskip}
L1 & 0.47 & 3.84 & 8.1 & 15.9/7 & $445_{-34}^{+26}$ & & 17.9/6 & $-1.12_{-0.12}^{+0.13}$ & $10.9_{-3.4}^{+8.5}$ & & 12.4/6 & $0.40_{-0.08}^{+0.14}$ & $8.6_{-2.1}^{+4.0}$\\[3pt]
L2 & 0.50 & 3.80 & 10.2 & 5.7/8 & $589_{-31}^{+27}$ & & 7.4/7 & $-1.00_{-0.10}^{+0.09}$ & $33.1_{-8.5}^{+14.4}$ & & 5.7/7 & $1.11_{-0.24}^{+0.26}$ & $7.9_{-1.9}^{+3.0}$\\[3pt]
L3 & 0.67 & 4.10 & 9.4 & 5.4/8 & $701_{-40}^{+33}$ & & 5.7/7 & $-0.87_{-0.11}^{+0.10}$ & $72.0_{-20.2}^{+30.2}$ & & 7.1/7 & $1.84_{-0.43}^{+0.64}$ & $6.2_{-1.8}^{+2.5}$\\[3pt]
\hline\noalign{\smallskip}
V1 & 0.53 & 3.89 & 5.2 & 11.6/7 & $352_{-44}^{+37}$ & & 8.4/6 & $-1.35_{-0.05}^{+0.17}$ & $3.8_{-1.0}^{+4.1}$ & & 7.7/6 & $0.21_{-0.05}^{+0.08}$ & $10.5_{-3.4}^{+4.8}$\\[3pt]
V2 & 0.44 & 3.82 & 10.9 & 15.4/8 & $576_{-31}^{+24}$ & & 15.4/7 & $-1.04_{-0.09}^{+0.08}$ & $31.2_{-8.1}^{+10.3}$ & & 10.0/7 & $1.11_{-0.21}^{+0.22}$ & $7.4_{-1.3}^{+2.3}$\\[3pt]
V3 & 0.62 & 3.90 & 12.7 & 10.5/8 & $788_{-32}^{+30}$ & & 12.0/7 & $-0.95_{-0.08}^{+0.07}$ & $72.8_{-15.3}^{+23.1}$ & & 10.3/7 & $2.38_{-0.41}^{+0.53}$ & $7.7_{-1.4}^{+2.1}$\\[3pt]
\hline\noalign{\smallskip}
R1 & 0.53 & 3.84 & 8.2 & 10.5/7 & $480_{-33}^{+30}$ & & 11.4/6 & $-0.95_{-0.11}^{+0.12}$ & $26.7_{-7.9}^{+16.3}$ & & 7.8/6 & $0.64_{-0.12}^{+0.22}$ & $5.7_{-1.3}^{+2.2}$\\[3pt]
R2 & 0.51 & 3.86 & 8.3 & 8.7/8 & $500_{-33}^{+31}$ & & 9.0/7 & $-1.08_{-0.11}^{+0.11}$ & $19.0_{-5.7}^{+10.7}$ & & 6.9/7 & $0.66_{-0.15}^{+0.18}$ & $8.3_{-2.2}^{+3.9}$\\[3pt]
R3 & 0.51 & 3.95 & 9.4 & 6.7/8 & $647_{-35}^{+34}$ & & 3.8/7 & $-1.13_{-0.07}^{+0.10}$ & $29.7_{-7.4}^{+12.9}$ & & 7.2/7 & $1.00_{-0.16}^{+0.35}$ & $10.1_{-2.2}^{+5.1}$\\
\noalign{\smallskip}\hline
\end{tabular}
    \begin{tablenotes}
      \small
      \item Columns are (1) name of the stack; (2) average redshift of the composite lens (weighted by the number of sources within 0.1-2 Mpc); (3) normalization factor of the second-order contribution to the estimator $\Delta\tilde{\Sigma}(r)$; (4) total signal-to-noise ratio within the range in radius used to fit the profile; (5) $\chi^2$ per degree of freedom for the best-fit SIS model; (6) best-fit SIS velocity dispersion; (7) $\chi^2$ per degree of freedom for the best-fit PLAW model; (8) best-fit PLAW slope; (9) best-fit PLAW normalization; (10) $\chi^2$ per degree of freedom for the best-fit NFW model; (11) best-fit NFW mass enclosed in the sphere of average contrast density $\Delta=200\rho_c$; (12) best-fit NFW concentration $c_{200}=R_{200}/r_s$. Best-fit parameters and $1\sigma$ uncertainties are obtained after marginalization over the 2nd free parameter for the PLAW and NFW models.
    \end{tablenotes}
  \end{threeparttable}
\end{table*}

\subsection{Combining weak and strong lensing}
To check the robustness of our results, in particular the determination of the NFW concentrations, we combined our weak-lensing measurements with the strong-lensing models obtained for eleven SARCS galaxy groups (see \citealt{sl2s, verdugo11,verdugo14}). These strong-lensing constraints were stacked as follows. Using Lenstool, we constructed mass maps for each of the eleven objects (SA22, SA39, SA50, SA63, SA66, SA72, SA80, SA83, SA112, SA123, and SA127) and we calculated the mass inside their respective Einstein radius. Then we combined these masses to obtain a \textit{mean mass} of $7.89\times10^{12}\mathrm{M_{\odot}}$. In order to be consistent with the weak-lensing analysis we set up $z_l = 0.5$ and $z_s = 1.5$, and we computed the corresponding Einstein radius for this mean mass, $\bar{\theta}_{E} = (5.4 _{-0.7}^{+0.3}) \arcsec$. The errors were estimated assuming that the main source of uncertainty comes from the lack of a precise measurement on the source redshift, i.e. $z_s = 1.5 \pm 0.5$ (we want to note that this value is probably over-estimated).\\
\indent To perform the combined weak+strong lensing fit, we simply added in the $\chi^2$ a constraint on this average Einstein Radius $\mathrm{R_E}=(32.8_{-4.3}^{+1.9})$ kpc, whose value is determined for the analytical mass models by solving numerically $g(\mathrm{R_E})=1$. For the weak-lensing constraints, we used the average profile for the stack S0, since the groups with a strong-lensing model do not fall in one single of the bins defined in Table 1. Because the strong-lensing signal allows us to reach a smaller radius, we checked the effect of including an extra central mass component (a point mass as described in Section 4.2). The results of the combined fit are given in Table 3.\\
First of all, we can see that, with the weak lensing only, adding a point mass does not change the best-fit parameters (third row compared to the first one). As shown in Section 4.2, we are not sensitive to the mass distribution within the inner regions of the groups, i.e. for $r<50$ kpc. The average shear profile only allows us to put an upper limit on the point mass, with values that are similar for the three mass models. Therefore, our approximation to exclude this mass component is valid.\\
\indent For the weak+strong lensing fit and without a point mass (second row), we obtained a significant change in the results for the PLAW and NFW models. Because the SIS mass model does not have a freedom on the shape of its profile, the fit is dominated by the weak-lensing constraints. Adding a single strong-lensing constraint does not lead to a significant change in the SIS velocity dispersion (agreement within the errors). On the other hand, the PLAW and NFW models are more sensitive to the value of the Einstein radius, which results in a more concentrated mass distribution (larger $c_{200}$ and steeper slope $\alpha$). As expected, adding the strong lensing-signal does not change significantly the total mass $M_{200}$. Interestingly, the combined weak+strong lensing presents a large likelihood ratio when compared to the weak-lensing only fit, with $\Delta\chi^2=8.5$ for the NFW model, and $\Delta\chi^2=13$ for the PLAW model: a single mass component no longer provides a good description of the total mass profile when the strong-lensing constraints are taken into account (especially for the PLAW model with a reduced $\chi^2/\mathrm{dof}\sim3$).\\
\indent If we add the central mass component in the model (forth row of Table 3), the quality of the fits is improved. As expected, the SIS velocity dispersion remains the same. For the PLAW and NFW models, we obtained best-fit parameters that are fully compatible (within their $1\sigma$ error bars) with the values derived from the weak-lensing only constraints: the two lensing regimes give consistent constraints on the total mass profile, providing the consideration of a supplementary mass component to the group-scale dark matter halo. While with the weak lensing only we could not constrain the value of this central mass, we managed to estimate it with the combined fit. Depending on the model chosen to describe the group-scale halo, we obtained a mass $M_0=(1.5-3.5)\times10^{12}\,\mathrm{M_{\odot}}$. The constraints on the point mass are rather loose, and adding this component in the fit increases as well the error bars on the other best-fit parameters. This result is not surprising because of the expected degeneracies between $M_0$ and the models' free parameters, i.e. increasing the central mass will require a less concentrated/steep profile for the group-scale halo. To quantify them, we estimated the Pearson correlation factor over the Monte Carlo drawings used in our fitting procedure. We obtained $r=0.68$ for $\alpha-M_0$ and $r=-0.76$ for $c_{200}-M_0$. The mass $M_{200}$ is mainly constrained by the weak lensing at large radii, and so has a very small correlation factor $r=-0.02$ with $M_0$.\\ 
\indent The interpretation of the point mass value is rather difficult because it is most likely resulting from the combination of different contributions: the baryonic mass of the central galaxy, its dark matter halo, the intra-group gas, or possible adiabatic contractions modifying the central shape of the group-scale dark matter halo. Given the upper limit on this central component, we derive a ratio $M_{200}/M_0>17$. \cite{okabe13} estimated a ratio $M_{200}/M_0>34$ from the stacked weak-lensing analysis of 30 clusters with $<M_{200}>\sim6\times10^{14}\mathrm{M_{\odot}}$. With their upper limit on $M_0$, \cite{umetsu14} found $M_{200}/M_0>30$ from the stacked analysis of 20 clusters with $<M_{200}>\sim13\times10^{14}\mathrm{M_{\odot}}$. Our result on the SARCS galaxy groups is comparable to those obtained for more massive galaxy clusters, which seems to indicate moderate variations of this ratio over the mass range of galaxy groups and clusters. \cite{han14} studied in detailed the stellar mass $M_\star$ associated to the central galaxy of group-scale haloes: for a halo mass $M_h\sim10^{14}\mathrm{M_{\odot}}$, the central stellar mass is expected to be $\sim5\times10^{11}\mathrm{M_{\odot}}$, i.e. a ratio $M_h/M_\star\sim200$ (see also \citealt{leauthaud12}, with expected ratios of 100-1000 for halo masses $10^{13}-10^{15}\mathrm{M_{\odot}}$). The ratio $M_{200}/M_0$ we derived is one order of magnitude smaller, which suggests that the value of the central mass component added to the total mass NFW profile cannot be only due to the stars within the central galaxy. A possible explanation could be a different central slope of the total NFW mass distribution. With a shallower density profile, or equivalently a core of almost constant density, the contribution of the dark matter halo would be larger in the inner part of the groups, thus requiring a less massive central point mass. The present lack of constraints to probe the central mass distribution of the composite SARCS lenses does not allow us to test such an hypothesis by using, for instance, a generalized or cored NFW profile. Studying the inner slope of the galaxy groups' mass profile is beyond the scope of this paper, and we refer the reader to the work presented in \cite{sand02,sand04}, where strong evidences of a shallower central dark matter profile are found for galaxy clusters.

\begin{table*}
\centering 
\label{table:fits}
\begin{threeparttable}
\caption{Results of the combined fit weak+strong lensing, with or without a central point mass.}
\begin{tabular}{l c c c c c c c c c c c c c}
\hline\hline\noalign{\smallskip}
  & \multicolumn{3}{c}{SIS} & & \multicolumn{4}{c}{PLAW} & & \multicolumn{4}{c}{NFW}\\
  \cline{2-4}\cline{6-9}\cline{11-14}
 Constraints & $\chi^{2}$/dof & $\left<\sigma_{v}\right>$ & $\left<M_{0}\right>$ & & $\chi^{2}$/dof & $\left<\alpha\right>$ & $\left<\Sigma_{0}\right>$ & $\left<M_{0}\right>$ & & $\chi^{2}$/dof & $\left<M_{200}\right>$ & $\left<c_{200}\right>$ & $\left<M_{0}\right>$\\
   &  & (km/s) & $10^{12}\mathrm{M_{\odot}}$ & &  & & $(\mathrm{M_{\odot}.pc^{-2}})$ & $10^{12}\mathrm{M_{\odot}}$ &  & & $(10^{14}\,\mathrm{M_{\odot}})$ & & $10^{12}\mathrm{M_{\odot}}$\\   
\noalign{\smallskip}\hline\noalign{\smallskip}
S0 & 15.8/8 & $516_{-19}^{+18}$ & - & & 13.1/7 & $-1.11_{-0.06}^{+0.07}$ & $21.0_{-4.1}^{+5.8}$ & - & & 6.8/7 & $0.73_{-0.10}^{+0.11}$ & $8.6_{-1.3}^{+2.1}$ & -\\[3pt]
S0+$\mathrm{R_E}$ & 18.6/9 & $531_{-14}^{+15}$ & - & & 26.1/8 & $-1.26_{-0.06}^{+0.05}$ & $12.6_{-2.9}^{+3.1}$ & - & & 15.3/8 & $0.64_{-0.10}^{+0.09}$ & $13.2_{-1.5}^{+2.7}$ & -\\[3pt]
\hline\noalign{\smallskip}
S0 & 16.2/7 & $504_{-34}^{+21}$ & $<2.33$ & & 14.3/6 & $-1.09_{-0.07}^{+0.06}$ & $20.6_{-4.0}^{+6.1}$ & $<0.96$ & & 6.9/6 & $0.72_{-0.10}^{+0.11}$ & $8.6_{-1.7}^{+2.0}$ & $<1.70$\\[3pt]
S0+$\mathrm{R_E}$ & 15.3/8 & $490_{-27}^{+28}$ & $1.51_{-0.85}^{+1.29}$ & & 22.2/7 & $-0.98_{-0.11}^{+0.09}$ & $23.5_{-5.9}^{+8.1}$ & $3.50_{-1.07}^{+1.31}$ & & 11.3/7 & $0.68_{-0.10}^{+0.12}$ & $5.8_{-1.3}^{+2.5}$ & $2.76_{-1.00}^{+1.24}$\\
\noalign{\smallskip}\hline
\end{tabular}
    \begin{tablenotes}
      \small
      \item Columns are (1) constraints used in the fit (S0 for weak lensing, $\mathrm{R_E}$ for strong lensing); (2) $\chi^2$ per degree of freedom for the best-fit SIS model; (3) best-fit SIS velocity dispersion; (4) central mass component added to the SIS model; (6) $\chi^2$ per degree of freedom for the best-fit PLAW model; (7) best-fit PLAW slope; (8) best-fit PLAW normalization; (9) central mass component added to the PLAW model; (10) $\chi^2$ per degree of freedom for the best-fit NFW model; (11) best-fit NFW spherical $M_{200}$; (12) best-fit NFW concentration $c_{200}$; (13) central mass component added to the NFW model. Best-fit parameters and $1\sigma$ uncertainties are obtained after marginalization over the other free parameters. The first two lines show the results of the fit without a point mass; the first line corresponds to the first line of Table 2. 
    \end{tablenotes}
  \end{threeparttable}
\end{table*}

\section{Mass-concentration relation of strong lenses}

\subsection{Concentrations of the SARCS galaxy groups}

 \begin{figure}
\includegraphics[width=\columnwidth, angle=0]{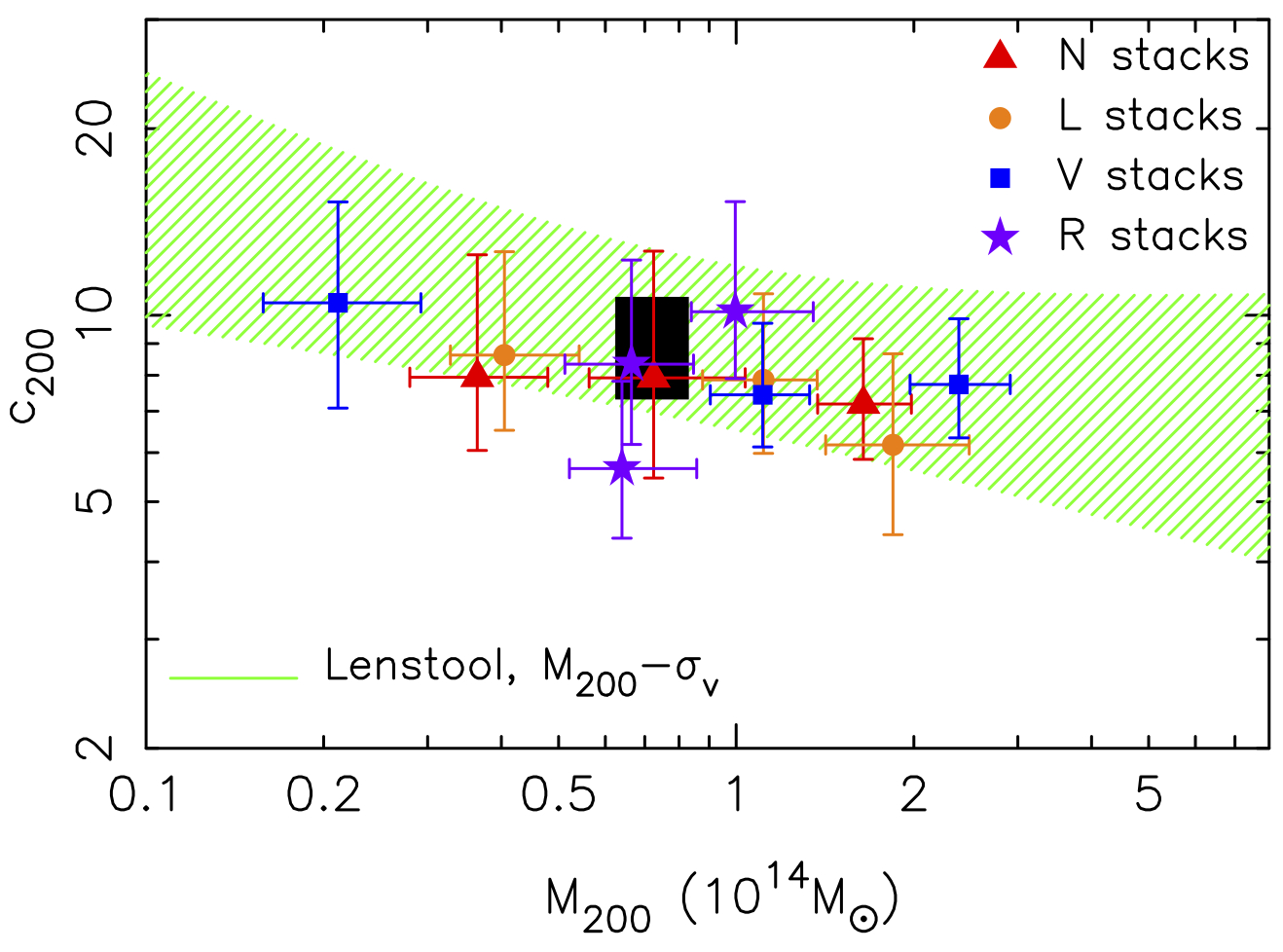}
\caption{Best-fit parameters of the NFW mass model for the stacks in richness $N$ (red triangles), luminosity $L$ (orange disks), SIS velocity dispersion $\sigma_v$ (blue squares), and arc radius $R_A$ (purple stars). The large black rectangle shows the $\left<M_{200}\right>\pm1\sigma$ and $\left<c_{200}\right>\pm1\sigma$ obtained using the 80 groups (stack S0). The green-hatched area delimits the $1\sigma$ uncertainty around the mass-concentration relation derived with Lenstool, using a $M_{200}(\sigma_v)$ scaling.}
\label{fig:concentration} 
\end{figure}

The results of the NFW model fitting are given Figure \ref{fig:concentration}, with similar concentrations for every stack. Using the 80 groups, we obtain a concentration $c_{200}=8.6_{-1.3}^{+2.1}$ for a mass $M_{200}=0.73_{-0.10}^{+0.11}\times10^{14}M_{\odot}$. Compared to the predictions from the numerical simulations of \cite{duffy08}, a halo with the same redshift and mass than the stack S0 should present a concentration of 3.5, i.e. we obtain an over-concentrated halo at the $\sim3\sigma$ level. Except for the stacks in arc radius (see below), increasing the stacking criterion gives a larger mass and a slightly lower concentration, hints of the expected $c(M)$ mass-concentration relation, which predicts less concentrated haloes of larger masses. We further tested this $c(M)$ relation with Lenstool, which uses a different approach that does not involve a stacking analysis.\\
\indent Lenstool ellipticity is defined as:
\begin{equation}
|\chi| = \frac{1-q^2}{1+q^2},
\end{equation}
where $q=b/a$ is the minor to major axis ratio. The ellipticity is expressed in a coordinate system in which $\chi_1$ is positive along the E-W axis, and $\chi_2$ is positive along the SE-NW diagonal. The conversion from the im2shape ellipticity $e=(a-b)/(a+b)$ is $\chi = 2e / (1 + e^2)$.\\
\indent In contrast to the stacking analysis described above, Lenstool estimates a scaling relation between an observable and the NFW mass $M_{200}$ (e.g. Equation \ref{eq:scale1}) in combination with the parameters of the $c(M)$ relation. The later follows the form of Equation \ref{eq:scale2}, with a scaling in redshift fixed at $(1+z)^{-0.45}$ \citep{duffy08}. Both relations are directly constrained from the measured ellipticities of the background galaxies (for the full sample of 80 lenses), assuming that the groups are modeled with 2D NFW potentials. For M weak lensing sources in our catalog, we define the likelihood as the product of M Gaussian likelihoods:
\begin{equation}
\mathcal{L} = \prod^{M}_i \frac{1}{\sqrt{2 \pi \sigma_{\chi_i}^2}}\exp \left( -\frac{1}{2} \frac{|\chi^s_i|^2}{\sigma_{\chi_i}^2} \right),
\end{equation}
where $|\chi^s_i|$ is the module of the predicted source ellipticity obtained by multiplying the amplification matrix $\mathcal{A}$ to the second brightness moments of each image $\mathcal{Q}$, through the equation $\mathcal{Q}^s_i = \mathcal{A} \mathcal{Q} \mathcal{A}^T$ \citep{bartelmann01}. In this formalism, the major and minor axes of a galaxy correspond to the eigenvalues of its $\mathcal{Q}$ matrix. Although it is not really needed for this work, this matrix transformation is valid both in the weak and in the strong lensing regime. In this work, we assume $\sigma_{\chi}^2 = \sigma_{\rm int}^2 + \sigma_{\rm meas_i}^2$, i.e. the variance is the quadratic sum of the intrinsic ellipticity and the shape measurement errors for each galaxy.\\
\indent With Lenstool, we can derive directly a $c(M)$ relation associated to a mass-observable scaling, and we tested this approach using either the optical richnesses or the SIS velocity dispersions, results summarized in Table 4. In both cases, we observe an anti-correlation between $M_{200}$ and $c_{200}$. The $c(M)$ relation derived with the scaling $M_{200}-\sigma_{v}$ has a (logarithmic) slope $B=0.07$ in fairly good agreement with predictions from numerical simulations, e.g. $B=0.084$ for the full sample of \cite{duffy08}. However, its normalization is much larger, and we obtain a very good match with the results from the stacking analysis (green-hatched area in Figure \ref{fig:concentration}). The scaling mass-richness leads to a $c(M)$ relation much steeper with $B=0.77$, result of a scaling shallower than expected, with $A=0.52$ instead of $\sim1$ (we have shown in Paper I that the large intrinsic dispersion of the groups leads to a reduced slope of the $\sigma_v-N$ scaling). Interestingly, for the $M_{200}-\sigma_{v}$ scaling, we obtain a lower limit on the slope $A-1\sigma=3.11$ in good agreement with its expected value $A=3$ (e.g. \citealt{evrard08,saro13}): the velocity dispersions derived in Paper I provide a tracer of the groups' mass that is less scattered than the optical observables. On the other hand, the scaling $M_{200}-N$ also leads to a $c(M)$ normalization that matches perfectly the average $c_{200}$ derived from the stacked analysis: no matter how the groups are combined or analyzed, the SARCS sample of strong lenses exhibits consistent concentrations that are larger than those measured in numerical simulations. A more precise analysis of the results derived with Lenstool via a combined fit of a scaling law and the $c(M)$ relation will be presented in a forthcoming dedicated paper.\\
\indent While an increase in richnesses, luminosities and SIS velocity dispersions translates into larger masses and slightly lower concentrations, we observe a different behavior for the stacks in arc radius. The stacks R1 and R2 present no significant change in the mass parameters $M_{200}$ and $\sigma_v$, but an increase in the concentration $c_{200}$, as well as a slightly steeper slope for the PLAW model. The stack R3 corresponds to a more massive composite lens with an increased $M_{200}$, but with a larger $c_{200}$. This correlation between the arc radius and the concentration might seem surprising given the $c(M)$ relation, and given that one would expect a correlation between the central projected mass (responsible of the size of the arc radius) and the total mass of a halo. Our results, which are consistent with the findings of \cite{oguri12}, suggest that the strong-lensing efficiency is mainly driven by the concentration of the haloes rather than their total mass. The scaling $M_{200}-\mathrm{R_A}$ appears to be weaker than the correlation $c_{200}-\mathrm{R_A}$, enough to outbalance the $c(M)$ relation. This correlation gives us a hint of the so-called strong-lensing bias: by combining lenses with a larger size of their gravitational arc (roughly equivalent to the Einstein radius), we introduce a selection bias resulting in a population with more concentrated projected mass distributions. Finally, we can explain the smaller value of the R1 concentration by a larger contribution of the central galaxy relative to the group-scale halo. For small arc radii, the mass of the central galaxy contributes enough to the lensing efficiency, and the projected mass distribution of the dark matter halo does not need to be very concentrated. Therefore, the concentration of groups with small arc radius derived by weak lensing is less biased towards high values.

\begin{table*}
\centering 
\label{table:fits}
\begin{threeparttable}
\caption{Lenstool constraints on the intermediate mass-observable scaling relation $M_{200}=M_{0}(X/X_{0})^{A}$, and the corresponding mass-concentration relation $c_{200}=c_0(M_{200}/M_{0})^{-B}(1+z)^{-0.45}$.}
\begin{tabular}{l c c c c c}
\hline\hline\noalign{\smallskip}
Scaling & $X_{0}$ & $M_0\,(10^{13}h^{-1}\mathrm{M_{\odot}})$ & $A$ & $c_0$ &$B$\\
\noalign{\smallskip}\hline\noalign{\smallskip}
$M_{200}-N$ & 20 & $3.87_{-0.95}^{+0.83}$ & $0.52_{-0.01}^{+0.25}$ & $10.8_{-3.4}^{+3.2}$ & $0.77_{-0.37}^{+0.02}$\\[3pt]
\hline\noalign{\smallskip}
$M_{200}-\sigma_v$ & 600 km/s & $5.29_{-0.89}^{+1.37}$ & $3.86_{-0.75}^{+0.09}$ & $7.8_{-1.6}^{+4.25}$ & $0.07_{-0.04}^{+0.26}$\\
\noalign{\smallskip}\hline
\end{tabular}
\begin{tablenotes}
\small
\item Columns are (1) mass-observable scaling used in the fit; (2) pivot to normalize the observable; (3) normalization of the scaling relation, which also corresponds to the pivot of the $c(M)$ relation; (4) logarithmic slope of the scaling relation; (5) normalization of the $c(M)$ relation; (6) logarithmic slope of the $c(M)$ relation. 
\end{tablenotes}
\end{threeparttable}
\end{table*}

\subsection{Evidence of a strong-lensing bias}

\begin{figure}
\includegraphics[width=\columnwidth, angle=0]{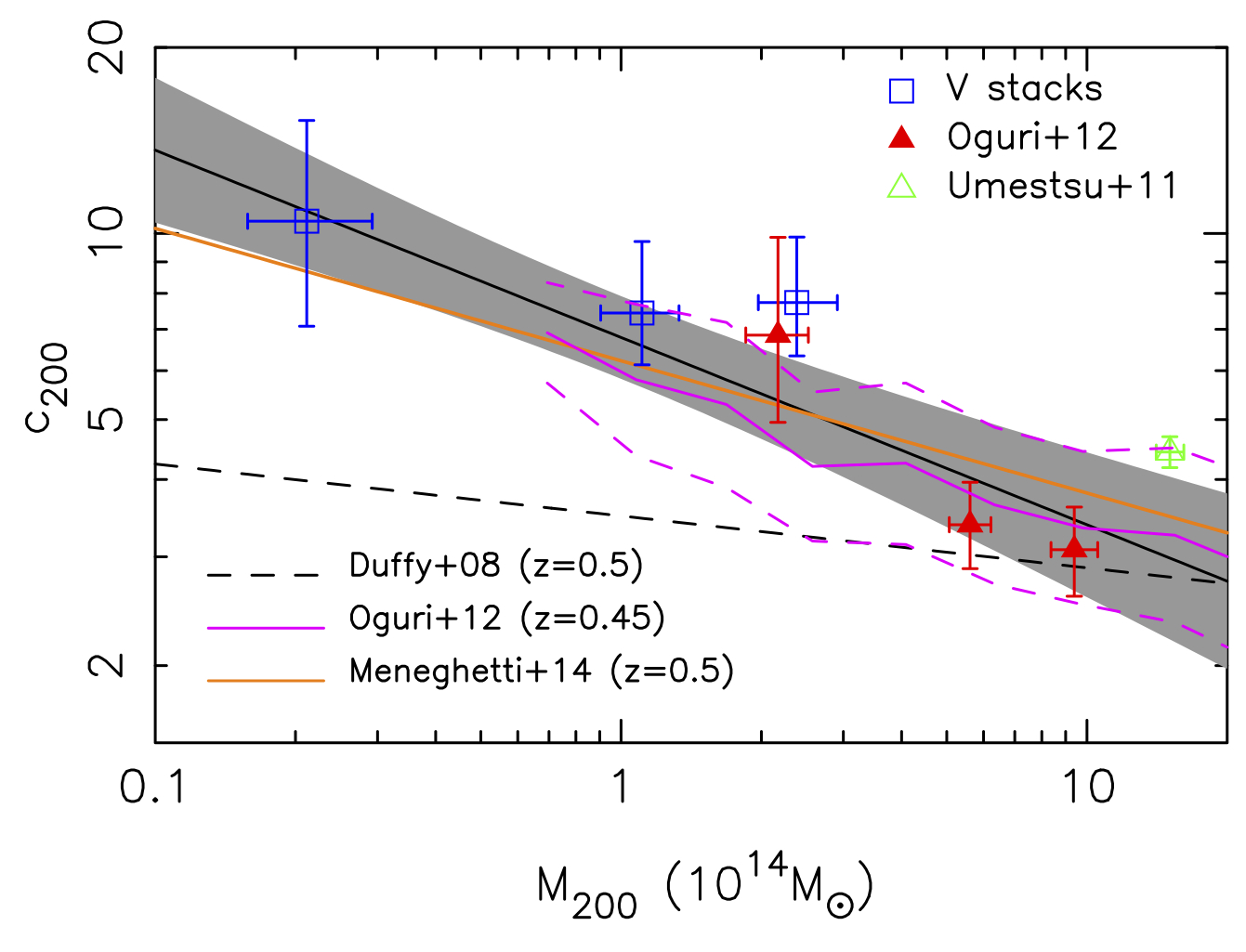}
\caption{Combined fit of the mass-concentration relation, using the results from the stacks in velocity dispersion (open-blue squares, weak lensing only) and the values derived by \cite{oguri12} (red triangles) and \cite{umetsu11b} (open-green triangle) from the stacked analysis of strong-lensing galaxy clusters. The best-fit relation was obtained assuming a scaling in redshift of $(1+z)^{-0.45}$, and the black solid line show the results for $z=0.5$ (the grey-shaded area delimits the statistical uncertainty from the best-fit parameters). Over-plotted are the relations derived from numerical simulations by \cite{duffy08} (black-dashed line, all haloes, $z=0.5$) and \cite{meneghetti14} (orange-solid line, strong-lensing selected haloes, $z=0.5$). The magenta-dashed lines show the $1\sigma$ limit around the average relation derived by \cite{oguri12} from semi-analytical predictions (lensing bias using weights from the arc cross section+Einstein radii, $z=0.45$).}
\label{fig:cMfit} 
\end{figure}

The NFW fit of the different stacks led to galaxy groups apparently over-concentrated compared to the expectations from numerical simulations for unbiased populations of dark matter haloes. Because the SARCS galaxy groups were selected from their strong-lensing signal, it is tempting to explain these large concentrations by a selection effect, i.e. strong lenses are a somehow biased population of haloes. To further test this strong-lensing bias, we combined our stacked galaxy groups with more massive strong lenses to fit the corresponding $c(M)$ relation. We used the 25 galaxy clusters binned in three stacks according to their Virial mass from \cite{oguri12}, and the stack of four massive galaxy clusters analyzed by \cite{umetsu11b}. These two studies made use of a strong+weak lensing analysis to derive NFW masses and concentrations, which, after conversion in our definition, leads to a range of nearly two decades in mass. The stacked clusters from \cite{umetsu11b} have an average redshift $z=0.32$, and the stacks from \cite{oguri12} have a redshift $z=0.46-0.48$, i.e. values similar to our average redshift $z=0.55$ for the full sample.\\
\indent Assuming a log-normal distribution for the NFW concentration parameter \citep{jing00}, we fitted a $c(M)$ relation expressed in logarithmic space:
\begin{equation}
\log\left(\frac{c_{200}}{c_{piv}}\right)=\log{c_0}-B\left(\frac{M_{200}}{M_{piv}}\right)-0.45(1+z),
\end{equation}
where the pivots $c_{piv}=5$ and $M_{piv}=10^{14}\,\mathrm{M_{\odot}}$ were chosen to be representative of the average mass and concentration of the combined samples, thus reducing the correlation in the best-fit normalization $c_0$ and slope $B$. We accounted for the slight differences in the average lens redshifts by rescaling them with a redshift evolution $(1+z)^{-0.45}$ \citep{duffy08}. To include in the fit error measurements on $M_{200}$ and $c_{200}$, and to allow for an intrinsic dispersion of the points around the best-fit relation, we used the BCES orthogonal estimator \citep{akritas96} in the same way as described in \cite{foex12}.\\
Because the V stacks (i.e. according to the SIS velocity dispersions) give the larger range in mass, we combined the corresponding masses and concentrations to those from \cite{oguri12} and \cite{umetsu11b}. The results of the BCES fit are the following: a slope $B=0.30\pm0.09$, a normalization $c_0=1.63\pm0.25$, and an intrinsic dispersion $\sigma_{\log{c_{200}}}=0.11$. Figure \ref{fig:cMfit} presents this best fit: the different samples of stacked strong lenses are well constrained by a single $c(M)$ relation over nearly two decades in mass. It has a slope much steeper than that obtained by \cite{duffy08}, and a normalization such as concentrations are higher over the mass range $10^{13}-10^{15}\,\mathrm{M_{\odot}}$, up to a factor $\sim3$ at the low-mass end.\\
\indent These results confirm the presence of a bias in the population of strong-lensing haloes. To explain why we obtain such discrepant concentrations compared to those from numerical simulations, one has to recall the differences in the way concentrations are estimated. From the simulation point of view, haloes are (usually) treated as spherical objects: the intrinsic shape is not accounted for, and the density contrasts are estimated by averaging within spheres, as for estimating masses and thus concentrations. In the case of our lensing analysis (as for that of \citealt{oguri12} and \citealt{umetsu11b}), we also assume spherical symmetry. However, the shear signal only probes the projected mass distribution. In other words, depending on its orientation with respect to the line of sight, a triaxial halo can have very different projected mass distributions, therefore different lensing signal. Intuitively, a prolate halo (cigar-shaped) with a major axis close to the line of sight will present an over-concentrated projected mass profile, leading to a larger 'lensing' concentration compared to the value that would be obtained by analyzing it in three dimensions and assuming spherical symmetry. It is, therefore, tempting to attribute the large concentrations of strong lenses to an orientation bias of triaxial haloes, rather than intrinsically over-concentrated objects.\\
\indent Several studies have explored this strong-lensing bias by extracting from numerical simulations only haloes with a large enough strong-lensing cross section. The analysis of such peculiar populations of haloes have led to $c(M)$ relations that exhibit a larger normalization and a steeper slope. In Figure \ref{fig:cMfit} we show, for instance, the recent work by \cite{meneghetti14} who simulated haloes that mimic the CLASH sample of strong-lensing galaxy clusters. By estimating the concentration of these haloes with their projected mass distribution, thus following what would be obtained from a lensing analysis, they derived a $c(M)$ relation in very good agreement with our results (slope of -0.21). \cite{oguri12} estimated the apparent concentrations of strong-lensing galaxy clusters using a semi-analytic approach (magenta lines in Figure \ref{fig:cMfit}), and they found a similar behavior: a strong-lensing bias resulting in larger concentrations. Their calculations fit very well the high-mass end of our $c(M)$ relation, with a similar slope, i.e. a larger increase of the concentrations for the lower mass systems.\\
\indent Because the strong lensing efficiency is related to the projected mass distribution of a halo, an intrinsically very massive object will most likely produce a strong lensing signal no matter its shape and orientation. In the case of small galaxy groups, large enough projected mass distribution can only be obtained for very elliptical haloes with a major axis close to the line of sight. Therefore, it is natural that a selection of haloes via strong lensing will result in a population more biased at lower mass scales. This mass-dependent selection bias translates into the observed $c(M)$ relation for strong lenses, with slopes much steeper than those obtained considering unbiased population of haloes. The study by \cite{giocoli14} using simulated clusters highlight this effect. They selected samples of haloes according to their Einstein radius and obtained steeper slopes for increasing $\theta_E$ (see their Table 2 and Figure 14), from 0.21 for haloes with $\theta_E>5''$ to 0.34 for haloes with $\theta_E>20''$: the larger the Einstein radius, the larger the central projected mass, which can only be achieved for low mass systems by a larger elongation, thus resulting in much larger apparent concentrations.\\
\indent Although apparent too large concentrations can be explained by a combination of projection effects and an orientation bias, one cannot simply rule out the possibility that strong lenses are intrinsically a biased population of objects: intrinsically more concentrated (in three dimensions) and/or intrinsically more elongated haloes. \cite{hennawi07} have studied the properties of haloes with a large strong-lensing cross section (see also \citealt{meneghetti10a}). They found that such haloes present a distribution of axis ratio that is very similar to that of 'normal' haloes and that the largest source of bias in the strong-lensing selection is an orientation bias. \cite{oguriblandford09} found however that in the case of very large Einstein radius, an additional bias in the shape of the haloes is required, with larger elongations to increase the lensing efficiency. \cite{hennawi07} found apparent concentrations (derived from the projected mass distribution) that are 34\% larger for the strong lenses, a result of the orientation bias+projection effects, mixed with intrinsically more concentrated haloes (in three dimensions). However, the later effect was found to be responsible of an increase in the apparent concentrations of only 18\%.\\
\indent Finally, we can mention here another possible bias in the population of strong lenses. The $c(M)$ relation obtained by \cite{prada12} highlighted an alternative to explain large concentrations: a very rapid accretion of matter resulting in haloes with a more compact configuration. Moreover, mergers are known to increase the strong-lensing cross-section \citep{zitrin12,redlich12}. Therefore, a population of strong lenses is likely to be biased towards both elongated haloes along the line of sight and compact configurations due to recent mergers. However, at the group-scale, it is less likely that the haloes accreted large fractions of their mass very recently, compared to more massive galaxy clusters. With our observational results, it is impossible to disentangle the different possible sources of concentration enhancement, though we have shown in Paper I that a non-negligible fraction of the systems present a complex light morphology, a possible sign of merging events. In the next Section, we provide however a simple way to derive a lower limit on the minor:major axis ratio required to match the observed concentrations with the predicted ones. The value we derived for the SARCS sample is not unrealistic, thus we cannot conclude on the necessity of an additional bias to simple projection effects.

\subsection{A toy model}
\begin{figure}
\includegraphics[width=\columnwidth, angle=0]{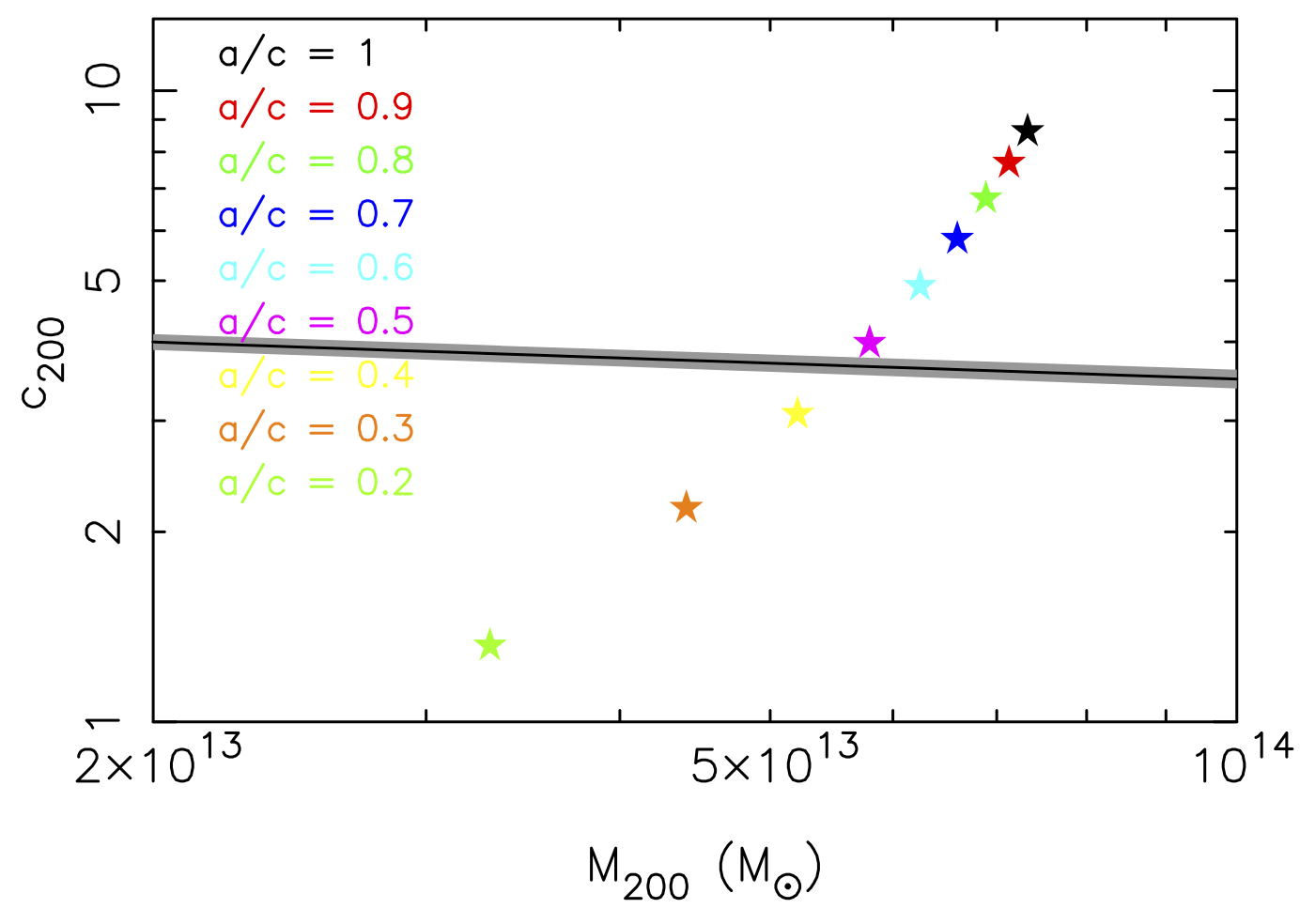}
\caption{Values of the spherical mass and concentration of a prolate NFW halo producing the same signal as that of a spherical NFW halo characterized by the S0 best-fit parameters. The different stars show the required $(M_{200,\,3D},c_{200,\,3D})$ as a function of the minor:major axis ratio $a/c$ (starting from $a/c=1$ in the top-right corner, and decreasing by 0.1). We consider here the case of a prolate halo ($a=b<c$) with a major axis aligned with the line-of-sight. The grey-shaded area and black curve indicate the relation of \citealt{duffy08} (all haloes, $z=0.5$).}
\label{fig:NFW_tri} 
\end{figure}

To reconcile the observed concentration of the SARCS sample with the predicted value from numerical simulations, let us consider the effect of the halo triaxiality on the observed lensing signal. The mass density of a triaxial NFW halo $\rho(R)$ is given by Equation \ref{eq:rhonfw}, with a radius $R$ expressed as \citep{jing02}:
\begin{equation}
R^2\equiv\frac{x^2}{a^2}+\frac{y^2}{b^2}+\frac{z^2}{c^2},\hspace{0.5cm}\left(a\leq b\leq c=1\right),
\end{equation}
where the coordinates $x$, $y$ and $z$ lie along the principal axes of the halo. $a$, $b$ and $c$ are the semi-minor, semi-intermediate and semi-major axes respectively of the iso-density ellipsoid defined by $R=1$. In the simplest case of a halo whose semi-major axis is oriented along the line of sight, one can show \citep{oguri03} that the convergence has the usual NFW functional form $\kappa(\zeta)$, with the projected radius
\begin{equation}
\zeta^2=\frac{x^2}{q_x^2}+\frac{y^2}{q_y^2},
\end{equation}
whose expressions for $q_x$ and $q_y$ are given in \cite{oguri03}. We can further simplify the problem by considering a prolate halo, i.e. $a=b<1$, a valid simplification when considering stacks of haloes whose projected ellipticity gets averaged. In doing so, we have $q_x=q_y=a/c$, leading to a density contrast $\Delta\Sigma$ that it the same as that of a spherical NFW halo (i.e. with spherical iso-densities) after a simple rescaling of the concentric distance $r\rightarrow\zeta=r/a$.\\
\indent The NFW profile $\Delta\Sigma(r)$ is a function of the ratio $r/r_s$, and its normalization is proportional to $(r_s\rho_0)$. Therefore, we can apply the rescaling $r_s\rightarrow r_s^{ell}=r_s/a$ and $\rho_0\rightarrow\rho_0^{ell}=a\rho_0$, so that the lensing signal produced by a prolate NFW halo with $(\rho_0^{ell},r_s^{ell})$ will be exactly the same as the signal of a spherical halo with $(\rho_0,r_s)$. To mimic the results derived from simulations, we can numerically integrate the density profile $\rho(R)$ of this prolate NFW halo $(\rho_0^{ell},r_s^{ell})$ in spheres, in order to find the 'spherical' radius $R_{200,\,3D}$ that defines the 'spherical' mass $M_{200,\,3D}=(800\pi/3)\rho_cR_{200,\,3D}^3$ for a given critical density $\rho_c$; the corresponding 'spherical' concentration is given by $c_{200,\,3D}=R_{200,\,3D}/r_s^{ell}$.\\
\indent This toy model provides a simple way to convert the 'lensing' masses and concentrations (i.e. derived from the shear signal using a spherical NFW halo) of a prolate halo into its 'spherical' values as they would be derived in numerical simulations (i.e. measuring masses in spheres rather than in iso-density ellipsoids). Figure \ref{fig:NFW_tri} presents the results of this toy model: as expected, the larger the elongation of the halo, the smaller the concentration and mass have to be to produce the same lensing signal. For the stack S0, we see that a couple $(M_{200,\,3D},c_{200,\,3D})$ in agreement with the prediction from \cite{duffy08} would have been obtained from the lensing analysis by using a prolate NFW halo with an axis ratio $a/c\sim0.5$, a value similar to the median elongation of dark matter haloes in numerical simulations (e.g. \citealt{hennawi07,giocoli14}). Because we considered here the case of a prolate halo perfectly aligned with the line of sight, our estimated axis ratio can only be interpreted as a lower limit: introducing an angle would reduce the lensing efficiency, which would require a larger ratio $a/c$ to produce the same shear profile (i.e. a mass distribution less stretched along the line of sight).\\
\indent If we assume that the signal we measured for the stack S0 is, indeed, produced by a prolate halo in agreement with the relation of \cite{duffy08}, we can use Figure \ref{fig:NFW_tri} to estimate the bias in our lensing measurements due to the hypothesis of spherical symmetry. With a $\Delta M_{200}\sim0.15\times10^{14}M_{\odot}$ and a $\Delta c_{200}\sim4.5$ (shifts between the pink and black stars), we get a mass overestimated by $\sim25\%$, and a factor $\sim2$ for the concentration. These values are in good agreement with the results derived from the weak-lensing analysis of simulated catalogs, with typical biases in mass of $\sim30-40\%$, and up to a factor 2 for the concentration in the case of highly prolate lenses (e.g. \citealt{corless07,corless08,feroz12}).

\section{Conclusions}
In this paper, we presented the results derived from the stacked weak-lensing analysis of a sample containing 80 strong-lensing galaxy groups. While in our first study (Paper I) the shear signal for the individual lenses was too noisy to derive reliable information on their mass distribution, we managed here to constrain the mass profiles of composite lenses. The stacked profiles were fitted by three mass models, the singular isothermal sphere, a power law mass distribution, and the classical NFW model. When combining the 80 lenses, we derived an average SIS velocity dispersion $\sigma_v=(516_{-19}^{+18})$ km/s. The best PLAW model is characterized by a normalization $\Sigma_0=(21.0_{-4.1}^{+5.8})\,\mathrm{M_{\odot}.pc^{-2}}$ and a slope $\alpha=(-1.11_{-0.06}^{+0.07})$, a value close to that of an isothermal mass distribution. For the NFW profile, we found $M_{200}=(0.73_{-0.10}^{+0.11})\times10^{14}\mathrm{M_{\odot}}$ and $c_{200}=(8.6_{-1.3}^{+2.1})$, a concentration in strong disagreement with predictions from numerical simulations, e.g. larger at the $3\sigma$ level compared to the prediction by \cite{duffy08}. Not only the best-fit PLAW was found to be consistent with the SIS model, but also with a likelihood ratio $\chi^2_{min,SIS}-\chi^2_{min,NFW}=9$, we concluded that for group-scale haloes, the isothermal mass distribution still provides a fairly good description of the total mass, compared to galaxy clusters presenting higher rejection levels of the SIS model.\\
\indent To check the reliability of our results derived from weak-lensing only, in particular the NFW concentration parameter, we combined the stacked shear profiles with an average Einstein radius, derived from eleven strong-lensing models constructed in previous papers. When introducing a central mass component in the total mass distribution (modeled by a simple point mass), we obtained results that are fully consistent with those from the weak-lensing only fitting. We derived an upper limit of the central mass $M_0$ leading to a ratio $M_{200}/M_0>17$, a value similar to those obtained for galaxy clusters. The comparison with the expected ratios between total mass and stellar mass within the central galaxy suggested the presence of another component or a modification of the classical NFW profile, for instance the presence of a core radius of constant density or a shallower inner slope of the density profile.\\
\indent When stacking the groups according to their richness, luminosity or SIS velocity dispersion, we obtained larger masses and smaller concentrations when increasing the selection threshold, hints of the expected mass-concentration relation. The opposite correlation was observed for the stacks in arc radius, with nearly no change in mass but a significant increase in $c_{200}$ for larger $\mathrm{R_A}$. We explained this behavior by a selection bias outbalancing the mass-concentration relation, i.e. larger strong-lensing efficiencies produced by more concentrated projected mass distributions rather than more massive haloes. The results of the stacked analysis were compared to those derived from a different approach, based on the Lenstool code. Rather than binning the lenses, Lenstool makes use of each galaxy group to constrain a scaling law to convert an observable into the NFW mass $M_{200}$ together with the mass-concentration relation. Using either the richness or the individual $\sigma_v$, we derived a $c(M)$ relation whose normalization is fully consistent with the concentrations obtained from the stacked analysis. The slope of the $M_{200}-\sigma_v$ scaling derived by Lenstool was found to be in good agreement with theoretical expectations, while the scaling $M_{200}-N$ is shallower than expected and leading to a too steep $c(M)$ relation.\\
\indent We combined our results with those derived from stacked analyses of strong-lensing galaxy clusters, and constrained the specific $c(M)$ relation of strong lenses over two decades in mass. We confirmed the existence of a strong-lensing bias: a $c(M)$ relation with a steeper slope and that predicts larger concentrations compared to what is found for unbiased populations of haloes. This mass-dependent enhancement of the concentrations has also been observed in numerical simulations, extracting haloes with large strong-lensing cross sections, and estimating concentrations from the projected mass distribution. Our $c(M)$ relation, extended to group-scale haloes, perfectly matches these predictions derived for galaxy clusters. Finally, we presented a toy model to derive a lower limit on the elongation of a prolate NFW halo in the case of perfect alignment of the major axis and the line of sight. We have shown that our average $(M_{200},c_{200})$ derived for a spherical NFW halo can be reconciled with the predictions of \cite{duffy08} assuming a prolate halo with a minor:major axis ratio $a/c\sim0.5$. Such an elongation being a realistic value, as observed in numerical simulations, we concluded that simple projection effects are sufficient to explain the apparently over-concentrated mass distributions of strong lenses.

\begin{acknowledgements}
G.F. acknowledges support from FONDECYT through grant 3120160.\\
V.M. acknowledges support from FONDECYT through grant 112074.\\
T.V. acknowledges support from CONACYT through grant 165365 and 203489 through the program Estancias posdoctorales y sab\'aticas al extranjero para la consolidaci\'on de grupos de investigaci\'on.\\
G.F, V.M., E.J., and M.L. acknowledge support from ECOS-CONICYT through grant C12U02.\\
The authors thank M. Oguri for providing the results of the semi-analytic model used to derive the mass-concentration relation of strong lenses.\\
Based on observations obtained with MegaPrime/MegaCam, a joint project of CFHT and CEA/DAPNIA, at the Canada-France-Hawaii Telescope (CFHT) which is operated by the National Research Council (NRC) of Canada, the Institut National des Science de l'Univers of the Centre National de la Recherche Scientifique (CNRS) of France, and the University of Hawaii. This work is based in part on data products produced at TERAPIX and the Canadian Astronomy Data Centre as part of the Canada-France-Hawaii Tele- scope Legacy Survey, a collaborative project of NRC and CNRS.
\end{acknowledgements}

%
%_____________________________________________________________

\bibliography{../references}

\section*{Appendix}
\counterwithin{figure}{section}
\setcounter{figure}{0}
\renewcommand{\thesubsection}{\Alph{subsection}}

\subsection{Profiles of mass-density contrast}
\begin{figure*}
\includegraphics[width=.9\paperwidth, angle=0]{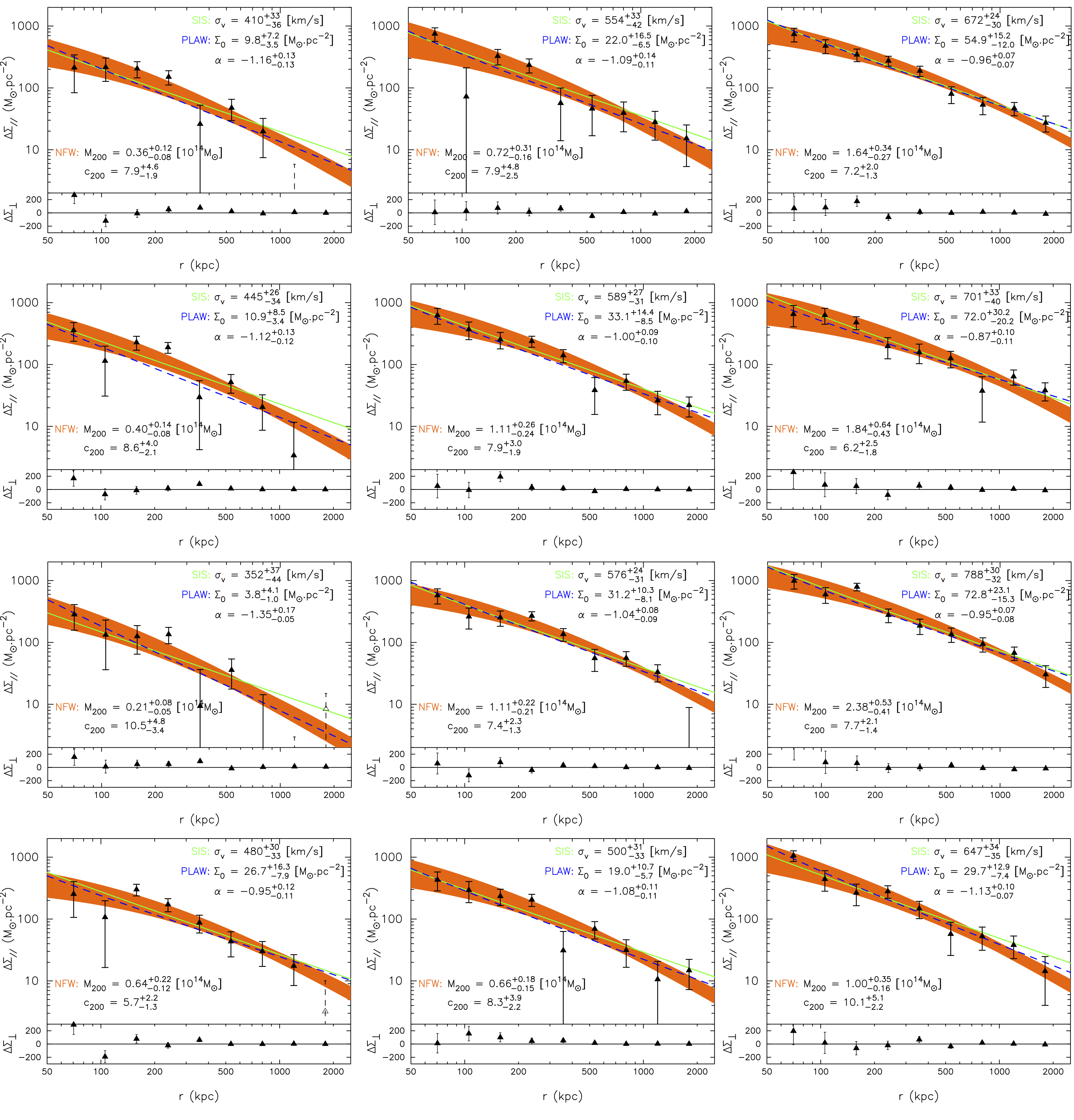}
\caption{Mass-density contrast for all the stacks in richness $N$ (first row), luminosity $L$ (second row), SIS velocity dispersion $\sigma_v$ (third row), and arc radius $R_A$ (last row). See Figure \ref{fig:stacks} for the legend.}
\label{fig:all} 
\end{figure*}

\counterwithin{figure}{subsection}
Figure \ref{fig:all} presents the profile for each stack, along with the best-fit SIS, PLAW, and NFW mass models. 

\end{document}